\documentclass[sigconf, nonacm]{acmart}

\usepackage[utf8]{inputenc}
\usepackage{graphicx}
\usepackage{xspace}
\usepackage{url}
\usepackage[inline]{enumitem}
\usepackage{amsmath}
\usepackage{tabularx}
\usepackage{xparse}
\usepackage{xcolor}
\usepackage{listings}
\usepackage{comment}
\usepackage{textcomp}
\usepackage{upquote}
\usepackage{caption}
\usepackage{subcaption}
\usepackage{tcolorbox}
\usepackage{float}
\usepackage[olditem,oldenum]{paralist}

\definecolor{delim}{RGB}{20,105,176}
\definecolor{numb}{RGB}{106, 109, 32}
\definecolor{string}{rgb}{0.64,0.08,0.08}

\lstdefinelanguage{json}{
    numbers=left,
    numberstyle=\small,
    frame=single,
    rulecolor=\color{black},
    showspaces=false,
    showtabs=false,
    breaklines=true,
    postbreak=\raisebox{0ex}[0ex][0ex]{\ensuremath{\color{gray}\hookrightarrow\space}},
    breakatwhitespace=true,
    basicstyle=\ttfamily\small,
    upquote=true,
    morestring=[b]",
    stringstyle=\color{string},
    literate=
     *{0}{{{\color{numb}0}}}{1}
      {1}{{{\color{numb}1}}}{1}
      {2}{{{\color{numb}2}}}{1}
      {3}{{{\color{numb}3}}}{1}
      {4}{{{\color{numb}4}}}{1}
      {5}{{{\color{numb}5}}}{1}
      {6}{{{\color{numb}6}}}{1}
      {7}{{{\color{numb}7}}}{1}
      {8}{{{\color{numb}8}}}{1}
      {9}{{{\color{numb}9}}}{1}
      {\{}{{{\color{delim}{\{}}}}{1}
      {\}}{{{\color{delim}{\}}}}}{1}
      {[}{{{\color{delim}{[}}}}{1}
      {]}{{{\color{delim}{]}}}}{1},
}

\newcommand{\definition}[1]{
   \noindent\textbf{\underline{#1}}
}
\newcommand{\emphasis}[1]{
   \noindent\textbf{\underline{#1}}
}

\newcommand\vldbdoi{XX.XX/XXX.XX}

\newcommand\vldbavailabilityurl{}
\begin{document}
\newcommand{\SERAPH}{Seraph\xspace}
\newcommand{\LANG}{Seraph\xspace}
\newcommand{\CYPHER}{Cypher\xspace}
\newcommand{\CQL}{CQL\xspace}
\newcommand{\RSPQL}{RSP-QL\xspace}
\newcommand{\SECRET}{SECRET\xspace}

% VALUES
\newcommand{\INTEGERS}{\ensuremath{\mathbb{Z}}\xspace}
\newcommand{\ALPHABETH}{\ensuremath{\Sigma}\xspace}
\newcommand{\STRINGS}{\ensuremath{\Sigma\ast}\xspace}
\newcommand{\TRUE}{\texttt{\textbf{true}}\xspace}
\newcommand{\FALSE}{\texttt{\textbf{false}}\xspace}
\newcommand{\NULL}{\texttt{\textbf{null}}\xspace}
\newcommand{\KEY}[1][]{\ensuremath{\ifx{#1} \empty k \else k_{#1} \fi} \xspace}
\NewDocumentCommand{\VALUE}{oo}{\ensuremath{v\IfValueT{#2}{{#2}}\IfValueT{#1}{_{#1}}}}
\newcommand{\LIST}{list\xspace}
\newcommand{\MAP}{map\xspace}
\newcommand{\PATH}{path\xspace}
\newcommand{\NODE}[1][]{\ensuremath{\ifx{#1} \empty n \else n_{#1} \fi} \xspace}
\newcommand{\RELATIONSHIP}[1][]{\ensuremath{\ifx{#1} \empty r \else r_{#1} \fi} \xspace}

% PROPERTY GRAPH
\newcommand{\LABELS}{$\mathcal{L}$\xspace}
\newcommand{\TYPES}{$\mathcal{T}$\xspace}
\newcommand{\PROPERTYKEYS}{\ensuremath{\mathcal{K}}\xspace}
\newcommand{\VALUES}{\ensuremath{\mathcal{V}}\xspace}
\newcommand{\NODEIDS}{\ensuremath{\mathcal{N}}\xspace}
\newcommand{\RELIDS}{\ensuremath{\mathcal{R}}\xspace}
\newcommand{\GRAPH}[1][]{\ensuremath{\ifx{#1} \empty G \else G_{#1} \fi} \xspace}
\newcommand{\NODES}[1][]{\ensuremath{\ifx{#1} \empty N \else N_{#1} \fi} \xspace}
\newcommand{\RELATIONSHIPS}[1][]{\ensuremath{\ifx{#1} \empty R \else R_{#1} \fi} \xspace}
\newcommand{\SOURCE}[1][]{\ensuremath{\ifx{#1} \empty src \else src_{#1} \fi} \xspace}
\newcommand{\TARGET}[1][]{\ensuremath{\ifx{#1} \empty trg \else trg_{#1} \fi} \xspace}
\newcommand{\PROPERTYFN}[1][]{\ensuremath{\ifx{#1} \empty \iota \else \iota_{#1} \fi} \xspace}
\newcommand{\LABELSFN}[1][]{\ensuremath{\ifx{#1} \empty \lambda \else \lambda_{#1} \fi} \xspace}
\newcommand{\TYPESFN}[1][]{\ensuremath{\ifx{#1} \empty \tau \else \tau_{#1} \fi} \xspace}

% TABLES
\newcommand{\NAMES}{\ensuremath{\mathcal{A}}\xspace}
\NewDocumentCommand{\NAME}{oo}{\ensuremath{a\IfValueT{#2}{{#2}}\IfValueT{#1}{_{#1}}}}

\newcommand{\RECORD}[1][]{\ensuremath{\ifx{#1} \empty u \else u{#1} \fi} \xspace}
\newcommand{\DOMAIN}{\ensuremath{dom}\xspace}
\NewDocumentCommand{\TABLE}{oo}{\ensuremath{T\IfValueT{#2}{^{#2}}\IfValueT{#1}{_{#1}}}}
\newcommand{\BAG}[1][]{\ensuremath{\ifx{#1} \empty B \else B_{#1} \fi} \xspace}
\newcommand{\BAGELEM}[1][]{\ensuremath{\ifx{#1} \empty b \else b_{#1} \fi} \xspace}

% TIME VARYING PROPERTY GRAPH
\newcommand{\TVPG}[1][]{\ensuremath{\ifx{#1} \empty \widetilde{G} \else  \widetilde{G}_{#1} \fi} \xspace}

% CATALOG
\newcommand{\GRAPHNAMES}{\ensuremath{\mathbf{N}}\xspace}

% TIME-BASED WINDOW
\newcommand{\TBW}[1][]{\ensuremath{\ifx{#1} \empty W \else  W_{#1} \fi} \xspace}
\newcommand{\OPENT}[1][]{\ensuremath{\ifx{#1} \empty o \else o_{#1} \fi} \xspace}
\newcommand{\CLOSET}[1][]{\ensuremath{\ifx{#1} \empty c \else c_{#1} \fi} \xspace}

% TIME-BASED SLIDING WINDOW
\newcommand{\WIDTH}[1][]{\ensuremath{\ifx{#1} \empty \alpha \else \alpha_{#1} \fi} \xspace}
\newcommand{\SLIDE}[1][]{\ensuremath{\ifx{#1} \empty \beta \else \beta_{#1} \fi} \xspace}
\newcommand{\WTIME}[1][]{\ensuremath{\ifx{#1} \empty t \else t_{#1} \fi} \xspace}
\newcommand{\TBSW}[1][]{\ensuremath{\ifx{#1} \empty \mathbf{W} \else  \mathbb{W}_{#1} \fi} \xspace}

% EVENT-BASED SLIDING WINDOW
\newcommand{\EBSW}[1][]{\ensuremath{\ifx{#1} \empty \mathbf{W} \else  \mathbb{W}_{#1} \fi} \xspace}

% PRESENT SUB-WINDOW
\newcommand{\PSW}[1][]{\ensuremath{\ifx{#1} \empty W^{'} \else  W_{#1}^{'} \fi} \xspace}

% TIME VARYING TABLES
\newcommand{\TVT}[1][]{\ensuremath{\ifx{#1} \empty \widetilde{T} \else  \widetilde{T}_{#1} \fi} \xspace}

\newcommand{\MATCH}{\texttt{\textbf{\textcolor{blue}{MATCH}}}\xspace}
\definecolor{keyword}{HTML}{2771a3}
\definecolor{pattern}{HTML}{b53c2f}
\definecolor{string}{HTML}{be681c}
\definecolor{relation}{HTML}{7e4894}
\definecolor{variable}{HTML}{107762}
\definecolor{comment}{HTML}{8d9094}

\lstset{
	numbers=none,
	stepnumber=1,
	numbersep=5pt,
	basicstyle=\small\ttfamily,
	keywordstyle=\color{keyword}\bfseries\ttfamily,
	commentstyle=\color{comment}\ttfamily,
	stringstyle=\color{string}\ttfamily,
	identifierstyle=,
	showstringspaces=false,
	aboveskip=3pt,
	belowskip=3pt,
	columns=flexible,
	keepspaces=true,
	breaklines=true,	
	captionpos=b,
	tabsize=2,
	frame=none,
    numbers=left, % show line numbers at the left
    numberstyle=\tiny\ttfamily, % style of the line numbers
}

\lstset{upquote=true}

\lstdefinelanguage{cypher}
{
	morekeywords={
		MATCH, OPTIONAL, WHERE, NOT, AND, OR, XOR, RETURN, DISTINCT, ORDER, BY, ASC, ASCENDING, DESC, DESCENDING, UNWIND, AS, UNION, WITH, ALL, CREATE, DELETE, DETACH, REMOVE, SET, MERGE, SET, SKIP, LIMIT, IN, CASE, WHEN, THEN, ELSE, END,
		INDEX, DROP, UNIQUE, CONSTRAINT, EXPLAIN, PROFILE, START, GRAPH, CONSTRUCT,
		REGISTER, QUERY, STREAM, RSTREAM, DSTREAM, ISTREAM, FROM, WINDOW, RANGE, SLICE, OUTPUT, EVERY, Events, Hours, Minutes, Seconds, COUNT,
		STARTING, Earliest, Latest, Event,
		EMIT, ON, EXIT, ENTERING, INTO, SNAPSHOT
	}
}

\newcommand{\mycdots}{\cdot\!\cdot\!\cdot}
\lstset{language=cypher,
	literate=*
	{...}{$\mycdots$}{1}
	{theta}{$\theta$}{1}
}

\newcommand{\code}[1]{
    \texttt{\textcolor{keyword}{\textbf{#1}}}
}
\title{Semantic Foundations of \SERAPH Continuous Graph Query Language}

%%
%% The "author" command and its associated commands are used to define the authors and their affiliations.
\author{Emanuele Falzone}
\email{emanuele.falzone@polimi.it}
\affiliation{%
  \institution{Politecnico di Milano}
%  \city{Milano}
%  \country{Italy}
}
\author{Riccardo Tommasini}
\email{riccardo.tommasini@ut.ee}
\affiliation{%
  \institution{University of Tartu}
%  \city{Tartu}
%  \country{Estonia}
}
\author{Emanuele Della Valle}
\email{emanuele.dellavalle@polimi.it}
\affiliation{%
  \institution{Politecnico di Milano}
%  \city{Milano}
%  \country{Italy}
}
\author{Petra Selmer}
\email{petra.selmer@neo4j.com}
\affiliation{%
  \institution{Neo4j}
%  \city{Milano}
%  \country{Italy}
}

\author{Stefan Plantikow}
\email{stefan.plantikow@neo4j.com}
\affiliation{%
  \institution{Neo4j}
%  \city{Milano}
%  \country{Italy}
}

\author{Hannes Voigt}
\email{hannes.voigt@neo4j.com}
\affiliation{%
  \institution{Neo4j}
%  \city{Milano}
%  \country{Italy}
}

\author{Keith Hare}
\email{keith.hare@neo4j.com}
\affiliation{%
  \institution{Neo4j}
%  \city{Milano}
%  \country{Italy}
}

\author{Ljubica Lazarevic}
\email{ljubica.lazarevic@neo4j.com}
\affiliation{%
  \institution{Neo4j}
%  \city{Milano}
%  \country{Italy}
}

\author{Tobias Lindaaker}
\email{tobias.lindaaker@neo4j.com}
\affiliation{%
  \institution{Neo4j}
%  \city{Milano}
%  \country{Italy}
}

\begin{abstract}
The scientific community has been studying graph data models for decades. Their high expressiveness and elasticity led the scientific community to design a variety of graph data models and graph query languages, and the practitioners to use them to model real-world cases and extract useful information. Recently, property graphs and, in particular, \CYPHER 9 (the first open version of the well-known Neo4j Inc.'s language) are gaining popularity. Practitioners find \CYPHER useful and applicable in many scenarios.
However, we are living in a streaming world where data continuously flows. A growing number of \CYPHER's users show interest in continuously querying graph data to act in a timely fashion. Indeed, \CYPHER lacks the features for dealing with streams of (graph) data and continuous query evaluation.
In this work, we propose \SERAPH, an extension of \CYPHER, as a first attempt to introduce streaming features in the context of property graph query languages. Specifically, we define \SERAPH semantics, we propose a first version of \SERAPH syntax, and we discuss the potential impacts from a user perspective. 
%The intent is to share with the scientific community our approach and gain inputs before prototyping a \SERAPH engine.
\end{abstract}

\maketitle

%\pagestyle{\vldbpagestyle}
%\begingroup\small\noindent\raggedright\textbf{PVLDB Reference Format:}\\
%\vldbauthors. \vldbtitle. PVLDB, \vldbvolume(\vldbissue): \vldbpages, \vldbyear.\\
%\href{https://doi.org/\vldbdoi}{doi:\vldbdoi}
%\endgroup
%\begingroup
%\renewcommand\thefootnote{}\footnote{\noindent
%This work is licensed under the Creative Commons %BY-NC-ND 4.0 International License. Visit %\url{https://creativecommons.org/licenses/by-nc-nd/4.0/} to view a copy of this license. For any use beyond those covered by this license, obtain permission by emailing \href{mailto:info@vldb.org}{info@vldb.org}. Copyright is held by the owner/author(s). Publication rights licensed to the VLDB Endowment. \\
%\raggedright Proceedings of the VLDB Endowment, Vol. \vldbvolume, No. \vldbissue\ %
%ISSN 2150-8097. \\
%\href{https://doi.org/\vldbdoi}{doi:\vldbdoi} \\
%}\addtocounter{footnote}{-1}\endgroup

%\ifdefempty{\vldbavailabilityurl}{}{
%\vspace{.3cm}
%\begingroup\small\noindent\raggedright\textbf{PVLDB %Artifact Availability:}\\
%The source code, data, and/or other artifacts have been made available at \url{\vldbavailabilityurl}.
%\endgroup

\section{Introduction}
% For many years, a lot of effort has been put into enhancing graph query languages, with several vendors providing different query languages~\cite{DBLP:journals/sigmod/Wood12, DBLP:conf/grades/RestHKMC16, DBLP:conf/sigmod/AnglesABBFGLPPS18,DBLP:conf/sigmod/FrancisGGLLMPRS18, DBLP:journals/tkde/SeoGL15, DBLP:conf/icpr/GiugnoS02, DBLP:journals/csur/AnglesG08}.
% \CYPHER~\cite{DBLP:conf/sigmod/FrancisGGLLMPRS18} is a language originally designed at Neo4j Inc.~\footnote{https://neo4j.com/} and first implemented by that company in 2011. 
% Since 2015, the openCypher\footnote{https://www.opencypher.org/} community has been making available open-source \CYPHER as well as many other resources to foster the rise of an ecosystem around property graphs and it is the best on-ramp to the graph query language standard\footnote{\url{https://www.gqlstandards.org/}} being developed by ISO. 

In the near future, graph processing systems and languages will have to address several challenges like the combination of different workloads, scalability beyond existing boundaries, and the need for new abstractions for querying and analysing~\cite{DBLP:journals/corr/abs-2012-06171}. In particular, we are living in a world where data continuously flows, and users demand to decide and act promptly to changes \cite{DBLP:journals/expert/ValleCHF09}. The growing popularity of sensors and smart-devices is pushing the boundaries of existing data systems. While real-time analytics is becoming central in data science projects, new languages are emerging to allow data scientists to express complex information needs and perform the analyses in real-time~\cite{DBLP:journals/sigmod/HirzelBBVSV18}. For instance, several streaming SQL extensions allow writing queries that compute every minute the average number of people crossing the road in the last 5 minutes. ~\cite{DBLP:conf/edbt/0001SVJ20}. However, such SQL extensions are inadequate to deal with the connected nature of many real-world scenarios, which, indeed, require graph data. Frequently, in those scenarios, Data Engineers have to craft highly complex ad-hoc solutions to perform continuous computations over streams of graphs. In practice, \textbf{a usable declarative language for graph stream processing is still missing}. 

In recent years, different vendors put a lot of effort into designing graph query languages~\cite{DBLP:journals/sigmod/Wood12, DBLP:conf/grades/RestHKMC16, DBLP:conf/sigmod/AnglesABBFGLPPS18,DBLP:conf/sigmod/FrancisGGLLMPRS18, DBLP:journals/tkde/SeoGL15, DBLP:conf/icpr/GiugnoS02, DBLP:journals/csur/AnglesG08}. In practice, the language ecosystem around property graphs is flourishing, and it is the best on-ramp to the graph query language (GQL) standard\footnote{\url{https://www.gqlstandards.org/}} being developed by ISO. Among the others,  Cypher~\cite{DBLP:conf/sigmod/FrancisGGLLMPRS18}, which was designed and implemented at Neo4j Inc.~\footnote{https://neo4j.com/} in 2011, proved to be useful and applicable in many scenarios \cite{cabra_2016, drakopoulos2016evaluating, hawes2015frappe, lysenko2016representing, robinson2013graph}. Moreover, it was open-sourced in 2015 under the name of OpenCypher\footnote{\url{https://www.opencypher.org/}} and is currently at the centre of the standardization process.
In this paper, we attack the problem of querying streams of property graphs starting from OpenCypher. In particular, we specify the syntax and the semantics of \SERAPH, a declarative language for the continuous query evaluation over streams of property graphs, aiming at answering the following research question:

% While it has been shown to be useful and applicable in many scenarios \cite{cabra_2016, drakopoulos2016evaluating, hawes2015frappe, lysenko2016representing, robinson2013graph}, it lacks the features for dealing with streams of graphs and continuous query evaluation.
% We are living in a world where data continuously flows and users are interested in continuously querying data to act in a timely manner \cite{DBLP:journals/expert/ValleCHF09}.
% Users want to exploit the temporal dimension of the data to extract relevant information, and continuously get updated as time passes.
% Such a goal is usually achieved by reducing the scope in time of the query applying a filter and periodically running it, e.g. compute every minute the average number of people crossing the road in the last 5 minutes.
% In order to fulfill these needs, starting from the early 2000s', the scientific community proposed \textit{continuous} query languages that natively support such features~\cite{DBLP:journals/vldb/ArasuBW06} and proposals also exist in the graph setting ~\cite{dell2014rsp, barbieri2009c,DBLP:conf/semweb/PhuocDPBEF12,DBLP:journals/ijswis/CalbimonteJCA12}.
% In this paper, we attack this problem by specifying the syntax and the semantics of a continuous property graph query language, aiming at answering the following research question:

\begin{center}
    \textit{Is it possible to compositionally extend the semantics and the syntax of \CYPHER to handle streams of property graphs and continuous queries?}
\end{center}

% Moving towards a positive answer, we introduce \SERAPH, a property graph query language that extends the core of \CYPHER to deal with streams of graphs and continuous query evaluation.
\noindent\textit{Outline.}
In Section~\ref{section:example}, we briefly illustrate a step-by-step running example, discussing the requirements that led to the formalization \SERAPH.
We provide an overview of the core of \CYPHER in Section~\ref{section:backgroud}, providing the preliminary knowledge needed to formalize \SERAPH.
The formal specification of the semantics of \SERAPH is given in Section~\ref{section:formal-specification} together with an initial version of its syntax. 
A system prototype, along with a performance evaluation, is presented in Section~\ref{section:system-prototype}.
Section~\ref{section:industrial-use-cases} illustrates real-world use cases.
Section~\ref{section:related-work} covers related works on other graph data models and graph query languages.
Future work and conclusions are given in Section~\ref{section:future-work} and Section~\ref{section:conclusions}, respectively.

%\begin{figure*}
%    \begin{subfigure}{.33\linewidth}
%        \centering
%        \includegraphics[width=0.75\linewidth]{content/images/administers.png}  
%        \caption{Allison gives Aaron a dose of Cortisol.}
%        \label{figure:example-administers}
%    \end{subfigure}
%        \begin{subfigure}{.33\linewidth}
%        \centering
%        \includegraphics[width=0.75\linewidth]{content/images/tested-for.png}  
%        \caption{Aaron tested positive for Norovirus.}
%        \label{figure:example-tested-for}
%    \end{subfigure}
%    \begin{subfigure}{.33\linewidth}
%        \centering
%        \includegraphics[width=0.75\linewidth]{content/images/is-with.png}  
%        \caption{Allison, House and Chase are in a meeting.}
%        \label{figure:example-is-with}
%    \end{subfigure}
%    \caption{Stream of graphs representing the events captured into the Kafka Hospital %Deployment.}
%    \label{figure:example-timeline}
%\end{figure*}

\begin{figure}
    \begin{subfigure}{0.3019191919\linewidth}
        \centering
        \includegraphics[width=\linewidth]{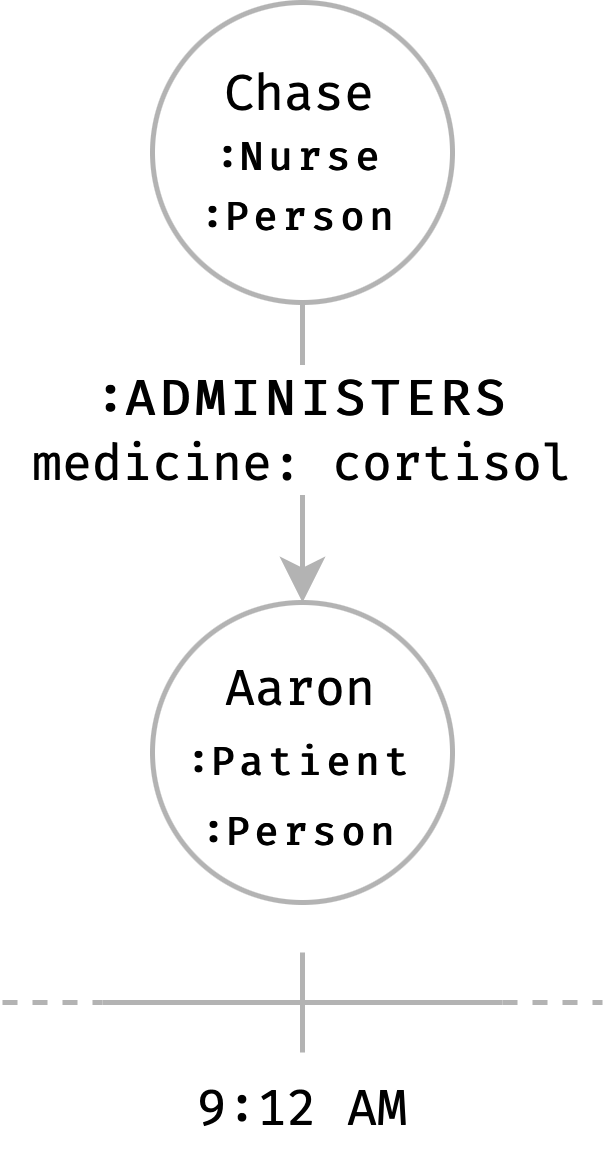}  
        %\caption{Allison gives Aaron a dose of Cortisol.}
        \caption{}
        \label{figure:example-administers}
    \end{subfigure}
        \begin{subfigure}{0.2524242424\linewidth}
        \centering
        \includegraphics[width=\linewidth]{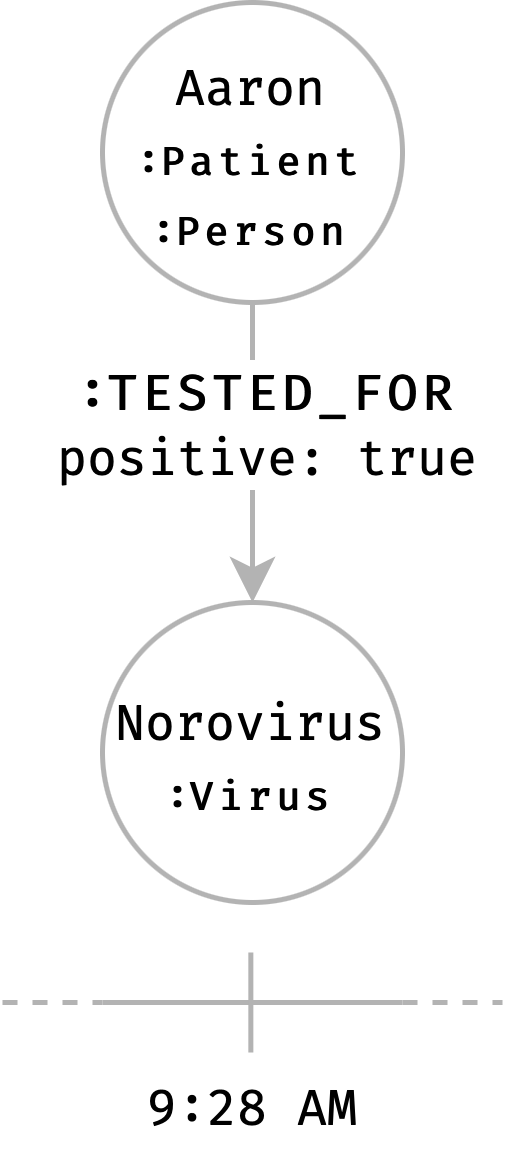}  
        %\caption{Aaron tested positive for Norovirus.}
        \caption{}
        \label{figure:example-tested-for}
    \end{subfigure}
    \begin{subfigure}{0.4256565657\linewidth}
        \centering
        \includegraphics[width=\linewidth]{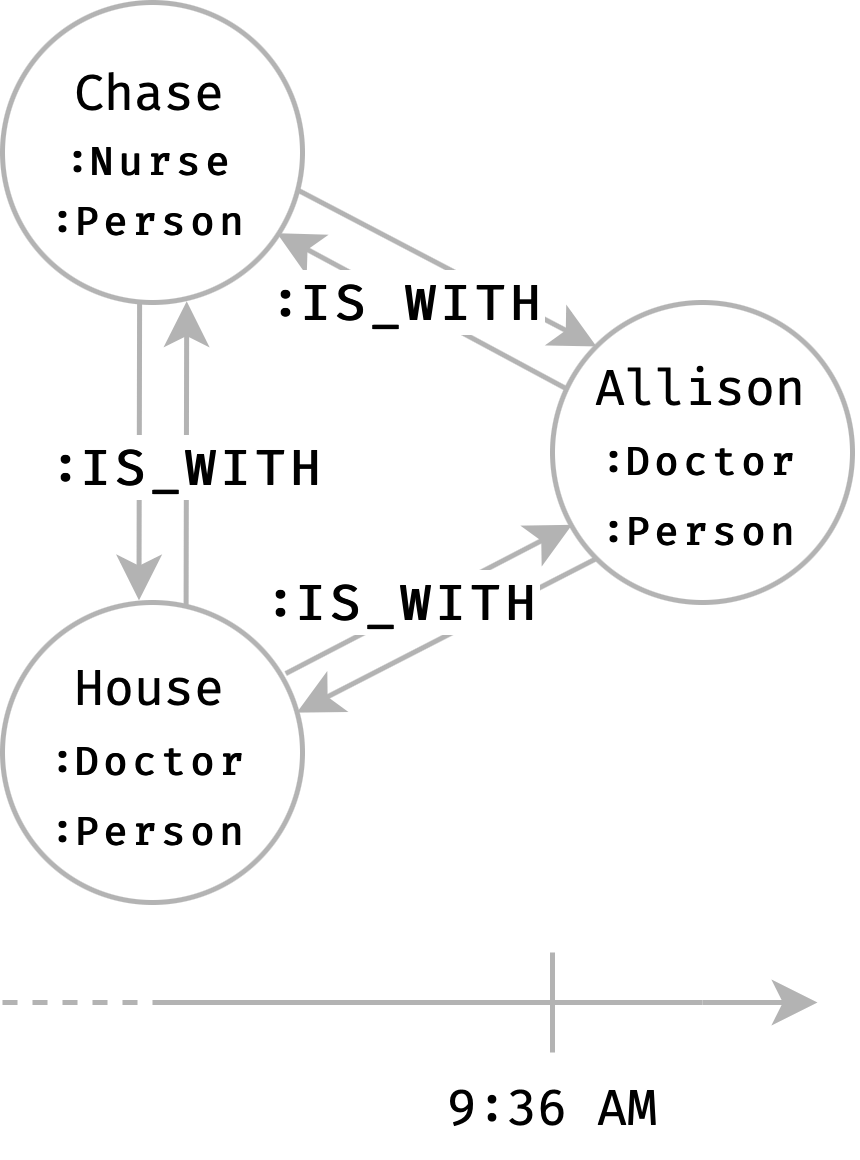}  
        %\caption{Allison, House and Chase are in a meeting.}
        \caption{}
        \label{figure:example-is-with}
    \end{subfigure}
    \caption{Stream of graphs representing the events captured into the Kafka Hospital Deployment.}
    \label{figure:example-timeline}
\end{figure}

\begin{figure}
\centering
\includegraphics[width=0.7121212121\linewidth]{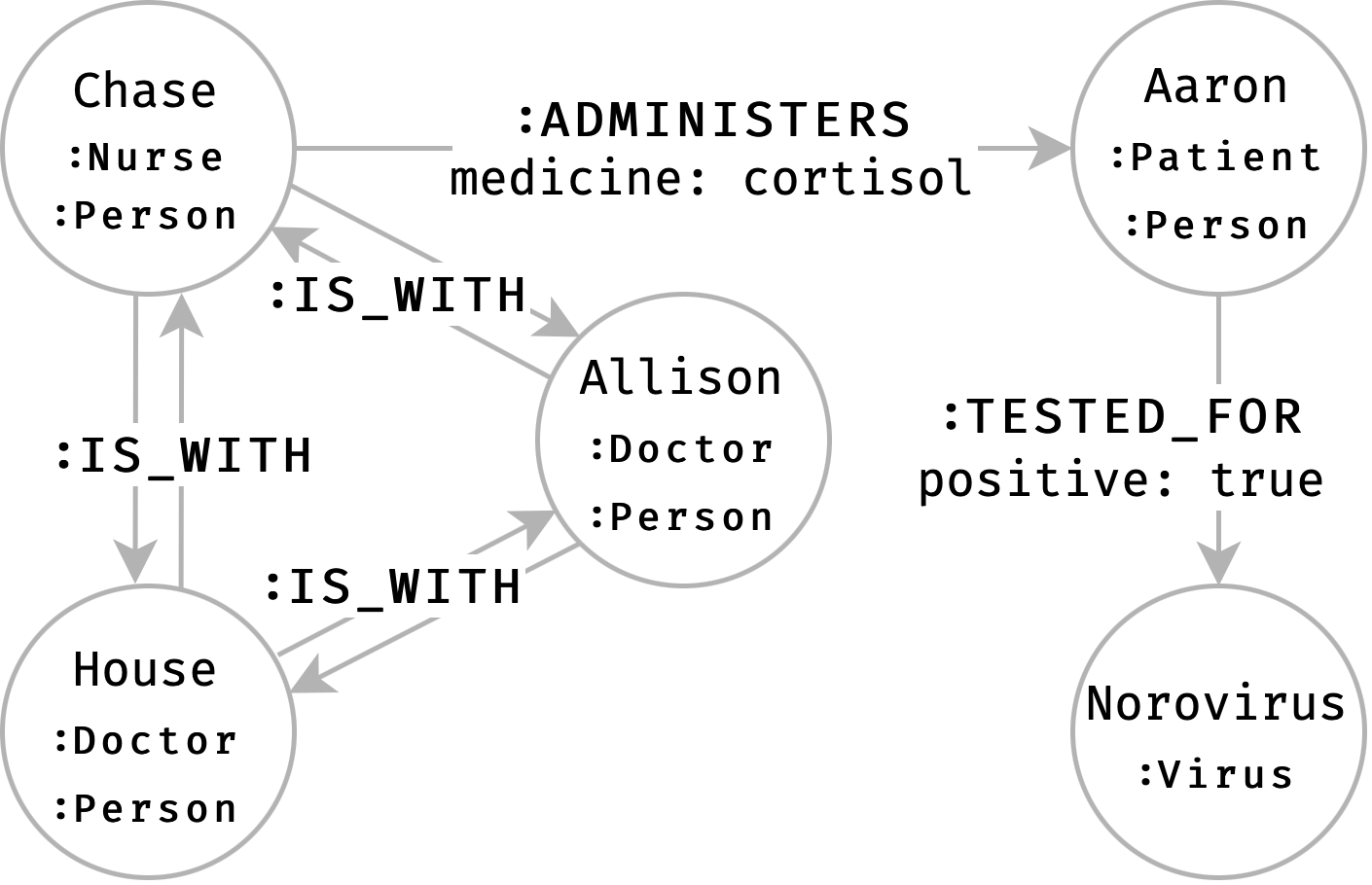}  
\caption{Graph resulting from loading of the events from 9:12 AM to 9:36 AM into a \texttt{Neo4j} instance.}
\label{figure:example-union}
\end{figure}

\section{\SERAPH by example}
\label{section:example}
In this section, we introduce a running example, discussing the language requirements that led to the formalization of the syntax and semantics of \SERAPH.

The Princeton–Plainsboro Teaching Hospital (PPTH) is known as one of the hospitals with the lowest mortality rate.
To protect both patients and hospital personnel, it has a series of strict policies to prevent the diffusion of viruses inside the buildings.
Such policies are continuously revised, and Dr. Lisa Cuddy, the hospital's head administrator, continuously searches for new tools and methodologies that can contribute towards her goal.

Recently, she realizes that a system able to track the contacts between infected patients and hospital personnel would be very efficient for prevention.
Indeed, identifying hospital operators that, in the last 4 hours, have been in contact with a patient that tested positive for a specific virus would be very useful: such operators would then be tested for the virus at the beginning of the next shift. 
Considering the high transmissibility and potential damage of viruses, Dr. Lisa Cuddy requires such information to be computed in real-time to notify the hospital operator about the possible infection.
Such operators are required to objectively evaluate the risk and adopt countermeasures to prevent the virus's diffusion.

PPTH uses a complex system, Galileo, that assists each hospital operator in daily routines and collects information about every event inside the hospital.
In particular, it stores every event in the form of property graphs inside the Kafka Hospital Deployment (KHD).

The timeline reported in Figure~\ref{figure:example-timeline} illustrates the events collected by Galileo from 9:12 AM to 9:36 AM.
Let us discuss the events in detail.

\begin{itemize}
\item \textbf{9:12 AM} As typically happens in hospitals, nurses administrate medicines to patients.
Chase, a recently employed nurse, gives Aaron a dose of Cortisol.
Figure~\ref{figure:example-administers} depicts the graph representing such an event.
\item \textbf{9:36 AM} Patients are periodically tested for viruses for both diagnosis and prevention purposes.
Aroon tested positive for a Norovirus.
Figure~\ref{figure:example-tested-for} depicts the graph representing such an event.
\item \textbf{9:28 AM} Allison, an experienced doctor, Dr. Gregory House, a genius diagnostician, and Chase are having a differential diagnosis meeting.
Figure~\ref{figure:example-is-with} depicts the graph representing such an event.
\end{itemize}

Unfortunately, Galileo is not able to track the contacts between infected patients and hospital personnel.
Consequently, Dr. Lisa Cuddy asks Elliot Alderson, the lead data scientist, to investigate a solution.
As a first step, Elliot asks Angela Moss, a member of the data engineering team, to load the events from 9:12 AM to 9:36 AM into a Neo4j instance, obtaining the graph reported in Figure~\ref{figure:example-union}.

Then, with the help of Dr. House, Elliot designed a \CYPHER query that retrieves the people to be tested for diseases.
Dr. House realized that it is not sufficient to retrieve the people that administer medicine to infected patients.
Due to the high transmissibility of the viruses, the colleagues, who have been in contact with the potentially infected hospital operator, have to be tested for the viruses.

\begin{figure}
\centering
\begin{lstlisting}[frame=single, escapechar=^]
MATCH (i:Patient)-[:TESTED_FOR {positive:true}]->(v:Virus),^\label{lst:cypher_query_match_patients}^
      (p:Person)-[:ADMINISTERS|IS_WITH*1..3]-(i)           ^\label{lst:cypher_query_match_people}^
RETURN DISTINCT p.name, v.name                             ^\label{lst:cypher_query_return}^
\end{lstlisting}
\caption{\CYPHER query to retrieve the hospital personnel to be tested for viruses.}
\label{figure:example-cypher-query}
\end{figure}

Consider the query reported in Figure~\ref{figure:example-cypher-query}. 
Line~\ref{lst:cypher_query_match_patients} exploits the \code{MATCH} clause to retrieve the patients \texttt{i} that tested positive for a virus \texttt{v}.
Line~\ref{lst:cypher_query_match_people} uses a variable-length path on \texttt{ADMINISTERS} and \texttt{IS\_WITH} relationships to match the people \texttt{p} to match the people that the infected patient \texttt{i} may have met.
Finally, Line~\ref{lst:cypher_query_return} uses the \code{RETURN} clause to project the name of the virus \texttt{v} and the name of the people \texttt{p} to be tested for such a virus.
Table~\ref{table:example-output-cypher} reports the result of the query evaluation.
As expected, both Chase, Allison, and House should be tested for the Norovirus.

\begin{table}
    \centering
    \texttt{
     \begin{tabularx}{\columnwidth}{X|X}
        \hline
        "Person" & "Virus" \\
        \hline
        "Chase" & "Norovirus" \\
        "House" & "Norovirus" \\
        "Allison" & "Norovirus" \\
        \hline
    \end{tabularx}
    }
    \caption{Results of the \CYPHER query evaluation.}
    \label{table:example-output-cypher}
\end{table}

Elliot is now provided with the query that computes the information need.
However, there are aspects of the data pipeline that Elliot cannot directly configure.
Indeed, the query has to be continuously evaluated on the events triggered in the last 4 hours.
Moreover, Elliot expects the results to be computed as soon as the underlying data is available.
In particular, he wants to get Chase returned at 9:28 AM and Allison and House at 9:36 AM.
Usually, Elliot relies on Angela to configure such aspects of the data pipeline.

However, Angela acts like a bottleneck: every time Elliot has a new information need, he has to wait for Angela to deploy and configure the data pipeline.
Elliot would like to have the ability to manage these aspects through the query language directly.
In this way, Elliot would no longer depend on Angela.
Moreover, Angela could spend all her time taking care of the management of the technological infrastructure. 

Elliot and Angela worked together on a solution.
They envision an expressive language that is provided with:

\begin{enumerate}[label=R\arabic*]
    \item \label{requirement:input} Precise control on data ingestion from the event stream.
    The language should offer operators that easily specify the input stream and the input stream's portion to consider.
    \item \label{requirement:evaluation} Precise control on evaluation time instant. The language should offer operators that intuitively allow for the specification of the evaluation time instants.
    \item \label{requirement:output} Precise control on output processing and emitting phase. The language should offer operators that allow for processing the output and for specifying the output stream where the result should be pushed.
\end{enumerate}

Moreover, to design a generic, easy to use solution, the language should fulfill the following requirements:

\begin{enumerate}[label=R\arabic*,start=4]
\item \label{requirement:ease-of-specification} Ease of specification. The language should allow for a declarative and easy specification of the information need.
\item \label{requirement:backwards-compatibility} Backwards compatibility. It should be possible to use an existing \CYPHER query to express the information need.
\item \label{requirement:generalization-and-specialization} Generalization and specialization. The language should allow a sufficient level of generality to be automatically applied for different data schema while allowing specialization on any given domain.
\end{enumerate}

\subsection{Solving the problem with \SERAPH}

\SERAPH aims at solving the problems discussed in the previous section. 
To this extent, \SERAPH builds upon the continuous-evaluation paradigm \cite{DBLP:journals/vldb/ArasuBW06}, which, as a consequence, influences the definition of \SERAPH query language.
We go in-depth into the semantics of \SERAPH in Section~\ref{section:formal-specification}, but here we provide some intuitions about it.
The sense of this choice is that the continuous evaluation can be viewed, at least when specifying the semantics, as a sequence of instantaneous evaluations~\cite{DBLP:conf/sigmod/TerryGNO92}. Therefore, given a fixed time instant, the operators can work in a time-agnostic way composing the semantics of \CYPHER in the one of \SERAPH.

The extension that we need to introduce is threefold:
\begin{itemize}
    \item Window operators. We need to let the \CYPHER operators process time-changing data, introducing window operators able to select a portion of the graphs contained in the stream (\ref{requirement:input},\ref{requirement:ease-of-specification},\ref{requirement:generalization-and-specialization}).
    \item Streaming operators. We need to work on the query's output to generate streams as output by introducing a new class of streaming operators that take as input time-varying tables and produce sequences of timestamped tables (\ref{requirement:output},\ref{requirement:ease-of-specification},\ref{requirement:generalization-and-specialization}).
    \item Continuous evaluation semantics. We need to extend the one-time \CYPHER semantics (\ref{requirement:backwards-compatibility}), defining the sequence of the evaluation time instants (\ref{requirement:evaluation}), and composing \SERAPH semantics with \CYPHER one, i.e., building \SERAPH evaluation around the evaluation process of \CYPHER (\ref{requirement:ease-of-specification},\ref{requirement:generalization-and-specialization}).
\end{itemize}

Elliot, exploiting Seraph capabilities, can compute the hospital personnel to be tested for viruses.
The \SERAPH query depicted in Figure~\ref{figure:example-seraph-query} continuously monitors the people to be tested.

\begin{figure}
\centering
\begin{lstlisting}[frame=single, escapechar=^]
REGISTER QUERY people_to_be_tested {              ^\label{lst:seraph_query_register}^
  FROM STREAM kafka://event-topic                 ^\label{lst:seraph_query_from_stream}^
  STARTING FROM Latest                            ^\label{lst:seraph_query_starting}^
  WITH WINDOW RANGE PT4H                         ^\label{lst:seraph_query_window}^
  MATCH
    (i:Patient)-[:TESTED_FOR {positive:true}]->(v:Virus),
    (p:Person)-[:ADMINISTERS|IS_WITH*1..3]-(i)
  RETURN DISTINCT p.name, 
                  v.name
  EMIT ON ENTERING EVERY 1 Event                  ^\label{lst:seraph_query_emit}^
  INTO kafka://result-topic                       ^\label{lst:seraph_query_into}^
}
\end{lstlisting}
\caption{Computing the people to be notified with \SERAPH.}
\label{figure:example-seraph-query}
\end{figure}

\begin{figure}
    \centering
    \includegraphics[width=0.5353535354\linewidth]{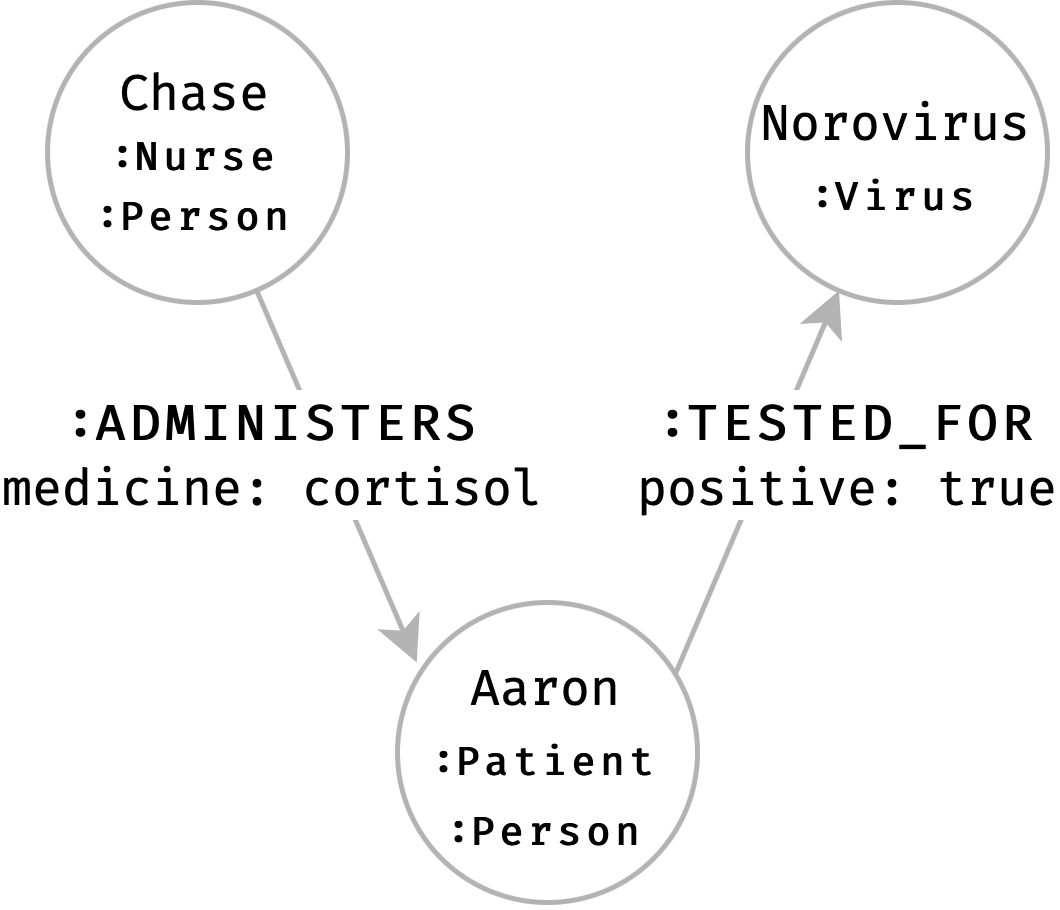}  
    \caption{Graphs resulting from the union of event occurred from 8:58 AM to 9:28 AM.}
    \label{figure:example-window-1}
\end{figure}
    
\begin{figure}
    \centering
    \includegraphics[width=0.7121212121\linewidth]{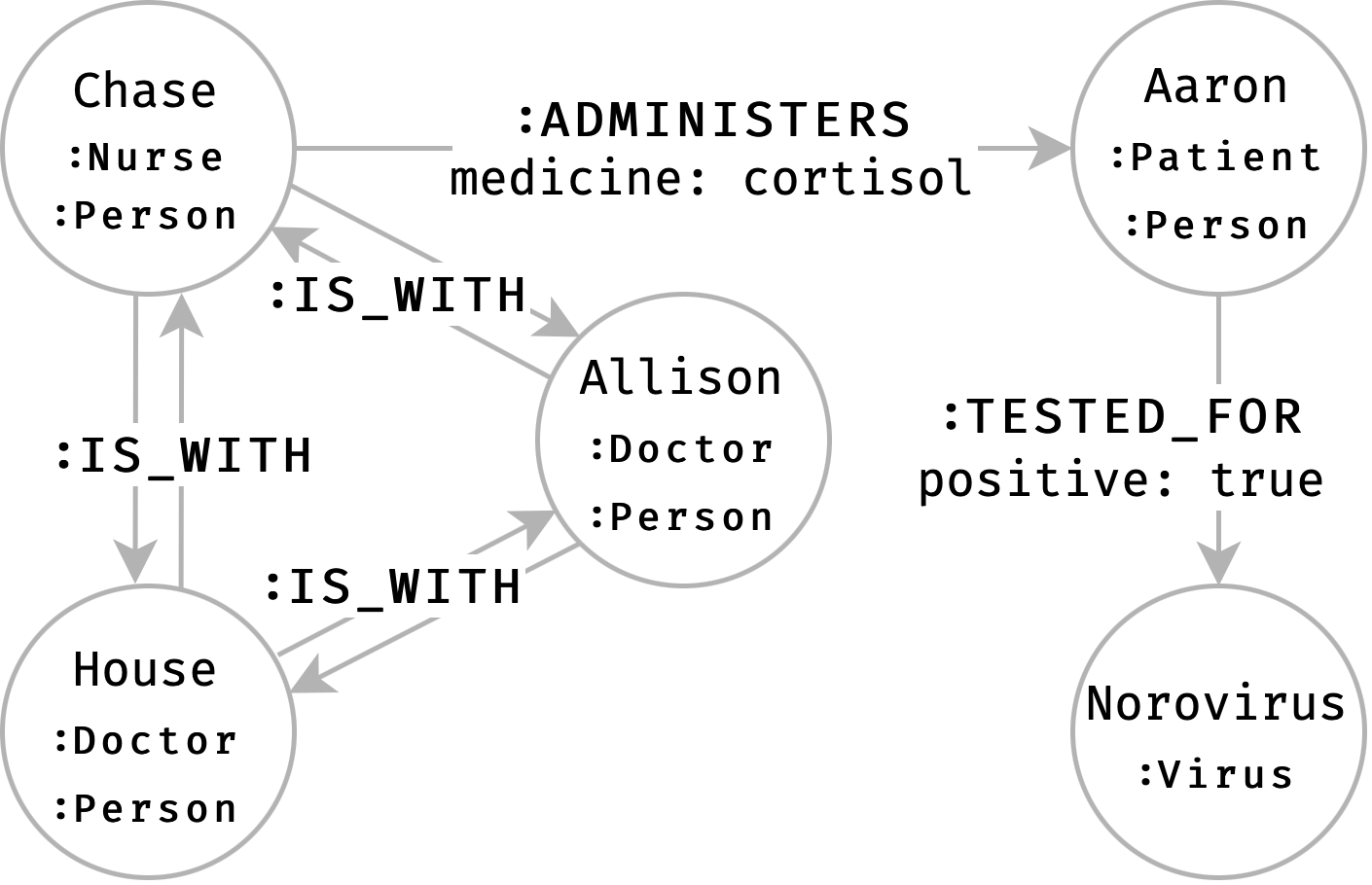}  
    \caption{Graphs resulting from the union of event occurred from 9:06 AM to 9:36 AM.}
    \label{figure:example-window-2}
\end{figure}

\begin{table}
    \centering
    \texttt{
    \begin{tabularx}{\columnwidth}{X|X}
        \hline
        "Person" & "Virus" \\
        \hline
        "Chase" & "Norovirus" \\
        \hline
    \end{tabularx}
    }
    \caption{Outputs of \SERAPH continuous query at 9:28 AM.}
    \label{table:example-output-1}
\end{table}

\begin{table}
    \centering
    \texttt{
    \begin{tabularx}{\columnwidth}{X|X}
        \hline
        "Person" & "Virus" \\
        \hline
        "Allison" & "Norovirus" \\
        "House" & "Norovirus" \\
        \hline
    \end{tabularx}
    }
    \caption{Outputs of \SERAPH continuous query at 9:36 AM.}
    \label{table:example-output-2}
\end{table}

Let us discuss the query.
The \code{REGISTER QUERY} clause at Line \ref{lst:seraph_query_register} allows for registering the query into the system that manages \SERAPH queries\footnote{The query handler permits to pause, restart, stop, start, and delete queries.}.
At Line \ref{lst:seraph_query_from_stream} the \code{FROM STREAM} clause is used to specify the source stream.
The \code{STARTING FROM} clause specifies the initial time instant that is used to define the sequence of the evaluation time instants.
In this case, \code{Latest} means that the first evaluation time instant is related to the last event processed.
The \code{WINDOW RANGE} operator at Line \ref{lst:seraph_query_window} sets up a 4 hours wide time-based sliding window.
The \code{MATCH} and \code{RETURN} clauses are the same as whose depicted in Figure~\ref{figure:example-cypher-query}.
Line \ref{lst:seraph_query_emit} specifies the \code{EMIT} clause.
In particular, the \code{ON ENTERING} operator restricts the reporting to new combinations of people and viruses so that each person is notified only once for the same virus.
The \code{EVERY} operator specifies the frequency of the evaluation process. 
Specifically, the code \code{EVERY 1 Event} tells that each new event triggers the evaluation process, thus notifying Chase at 9:28 AM and Allison and House at 9:36 AM, in the example. 
At last, the \code{INTO} clause is used to specify the target destination.

To summarize, every time a new event arrives, the system collects the events that occurred in the last 4 hours under a unique graph, and it queries such a graph to retrieve the people who have been in contact with an infected patient.
The resulting stream, which consists of tables, is pushed into a Kafka topic.

Let us analyze the output of the query at different time instants.
\begin{itemize}
    \item \textbf{9:12 AM} The only event captured is the one depicted in Figure~\ref{figure:example-administers}.
    No patients have been tested positive. Thus no hospital personnel should be notified.
    \item \textbf{9:13-9:27 AM} The query emits no event.
    \item \textbf{9:28 AM} The graphs in Figure~\ref{figure:example-administers} and Figure \ref{figure:example-tested-for} are merged as shown in Figure~\ref{figure:example-window-1}.
    Such an operation is automatically performed by \SERAPH under the unique name assumption \cite{tao2010integrity}.
    The graph is then queried to retrieve the people to be tested.
    Chase has to be notified of the potential exposure to Norovirus.
    The output of the query is depicted in Table~\ref{table:example-output-1}.
    \item \textbf{9:29-9:35 AM} The query emits no event.
    \item \textbf{9:36 AM} The graphs in Figure~\ref{figure:example-timeline} are unified as shown in Figure~\ref{figure:example-window-2}.
    Such a graph is then queried to retrieve the people to be tested for viruses.
    However, since we do not want to notify the same people for the same virus twice, only new combinations of person and virus w.r.t. the previous evaluation time instant, are reported.
    As expected, only Allison and House have to be notified of the potential exposure to Norovirus.
    The output of the query is depicted in Table~\ref{table:example-output-2}.
\end{itemize}

\begin{figure}
\centering
\includegraphics[width=0.7803030303\linewidth]{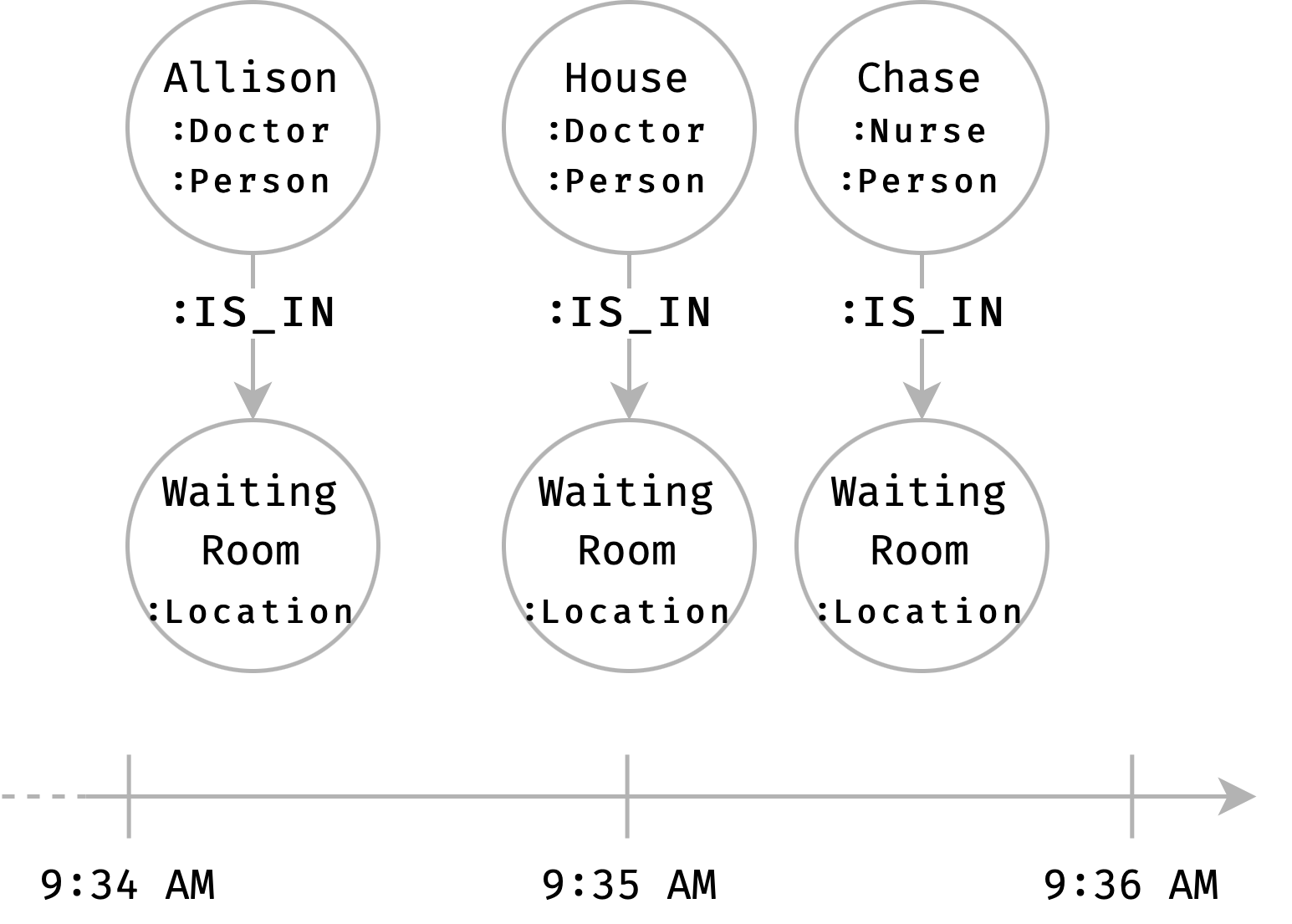}  
\caption{Localization events whenever a person is in a room.}
\label{figure:example-is-in}
\end{figure}

\emphasis{Producing a stream of graphs.}
The query depicted in Figure~\ref{figure:example-seraph-query} produces a stream of tables.
However, one can be interested in producing a stream of graphs instead of a stream of tables.
This allows for re-using the produced stream as input for other \SERAPH queries.

Doctors and nurses are equipped with a Smart Identification Tag (SIT).
Besides displaying information about the hospital operator, such a tag integrates a small Beacon that emits Bluetooth Low Energy signals.
Rooms, corridors, and stairs are provided with antennas that, exploiting the signals sent by the SITs, allows for the identification of the positions of people inside the hospital.
Each antenna is responsible for triggering localization events whenever a person is in a room.
An example of localization events is depicted in Figure~\ref{figure:example-is-in}.
The query produces a stream of events that whenever two or more people are in the same room is depicted in Figure~\ref{figure:example-seraph-query-output-stream-graph}.

Let us discuss the query.
The \code{REGISTER QUERY} clause at Line \ref{lst:seraph_graph_register_query} allows for registering the query.
At Line \ref{lst:seraph_graph_from_stream} the \code{FROM STREAM} clause is used to specify the source stream, which is the same as the one in the previous query.
The \code{STARTING FROM Latest} clause specifies the initial time instant that is used to define the sequence of the evaluation time instants.
The \code{WINDOW RANGE} operator at Line \ref{lst:seraph_graph_window} sets up a 2 minutes wide time-based sliding window.
The \code{MATCH} clause retrieves all the pairs of people in the same room.
The code \code{CONSTRUCT CREATE} \dots \code{RETURN GRAPH} clause at Line~\ref{lst:seraph_graph_return_graph} specifies that a new stream of graphs should be returned.
In particular, the \code{CREATE} statement at Line~\ref{lst:seraph_graph_construct} allows for creating new \texttt{IS\_WITH} relationships connecting two people that happened to be in the same room in 2 minutes time range.
At Line \ref{lst:seraph_graph_emit} the \code{EMIT ON ENTERING} operator allows for obtaining only the elements entering into the window.
The \code{EVERY} operator specifies the frequency of the evaluation process.
In this case, a minute is specified using \texttt{PT1M}.
The \code{INTO} clause then specifies the target destination of the stream, which is the same used as the input stream in the query depicted in Figure~\ref{figure:example-seraph-query}.

In this section, we briefly showcased the capabilities of \SERAPH with the aid of a running example.
In Section~\ref{section:industrial-use-cases}, real-world use cases are shown to highlight the industrial exploitability of the introduced language.

\begin{figure}[t]
\centering
\begin{lstlisting}[frame=single, escapechar=^]
REGISTER QUERY people_in_the_same_location {^\label{lst:seraph_graph_register_query}^
  FROM STREAM kafka://room-topic   ^\label{lst:seraph_graph_from_stream}^
  STARTING FROM Latest             ^\label{lst:seraph_graph_starting_from}^
  WITH WINDOW RANGE PT2M           ^\label{lst:seraph_graph_window}^
  MATCH
    (p1:Person)-[:IN]->(room:Room),
    (p2:Person)-[:IN]->(room)
  CONSTRUCT CREATE                 ^\label{lst:seraph_graph_construct}^
    (p1)-[:IS_WITH]->(p2)
  RETURN GRAPH                     ^\label{lst:seraph_graph_return_graph}^
  EMIT ON ENTERING EVERY PT1M      ^\label{lst:seraph_graph_emit}^
  INTO kafka://event-topic         ^\label{lst:seraph_graph_into}^
}
\end{lstlisting}
\caption{\SERAPH query to compute the people in the same room.}
\label{figure:example-seraph-query-output-stream-graph}
\end{figure}

\section{Background: the \CYPHER language}
\label{section:backgroud}

In the following, we provide the formal specification of the core of \CYPHER, which is a declarative query language for property graphs. The core of \CYPHER consists of: 
\begin{itemize}
    \item a data model that includes values, graphs, and tables;
    \item a query language that includes expressions, patterns, clauses, and queries.
\end{itemize}

Section \ref{section:backgroud:data-model} discusses the data model, while Section \ref{section:backgroud:evaluation-semantic} defines the evaluation semantic of the query language.
For brevity, we only report the definitions that we need to formally define \LANG.
However, we invite the interested reader to consult~\cite{DBLP:conf/sigmod/FrancisGGLLMPRS18} for the whole \CYPHER formalization.

%%%%%%%%%%%%%%%%%%%%%%%%%%%%%%%%%%%%%%%%%%%%%%%%%%%%%%%%%%%%%%%%%%%%%%%%%%%%%%%%%%%%%
%                               Data model                                          %
%%%%%%%%%%%%%%%%%%%%%%%%%%%%%%%%%%%%%%%%%%%%%%%%%%%%%%%%%%%%%%%%%%%%%%%%%%%%%%%%%%%%%
\subsection{Data model}\label{section:backgroud:data-model}

\definition{Values.}
We consider three disjoint sets \PROPERTYKEYS of property keys, \NODEIDS of node identifiers, and \RELIDS of relationship identifiers. 
These sets are all assumed to be countably infinite. 
For this presentation of the model, we assume two base types: the integers \INTEGERS, and the type of finite strings over a finite alphabet \ALPHABETH. 

The set \VALUES of values is inductively defined as follows:
\begin{itemize}
    \item Identifiers, i.e. elements of \NODEIDS and \RELIDS, are values
    \item Base types (elements of \INTEGERS and \STRINGS) are values
    \item \TRUE, \FALSE and \NULL are values
    \item $\LIST()$ is a value (empty list), and if $\VALUE[1], \VALUE[2], \dots , \VALUE[m]$ are values, for $m > 0$, then $\LIST(\VALUE[1], \VALUE[2], \dots , \VALUE[m])$ is a value
    \item $\MAP()$ is a value (empty map), and if $\KEY[1], \KEY[2], \dots , \KEY[m]$ are distinct property keys and $\VALUE[1], \VALUE[2], \dots , \VALUE[m]$ are values, for $m > 0$, then $\MAP((\KEY[1],\VALUE[1]), (\KEY[2],\VALUE[2]), \dots , (\KEY[m],\VALUE[m]))$) is a value
    \item If \NODE is a node identifier, then $\PATH(\NODE)$ is a value. 
    If $\NODE[1], \NODE[2], \dots , \NODE[m]$ are node ids and $\RELATIONSHIP[1], \RELATIONSHIP[2], \dots , \RELATIONSHIP[m-1]$ are relationship ids, for $m > 1$, then $\PATH(\NODE[1], \RELATIONSHIP[1], \NODE[2], \RELATIONSHIP[2], \dots , \NODE[m-1], \RELATIONSHIP[m-1], \NODE[m])$ is a value. 
\end{itemize}

\definition{Property graph.}

Let \LABELS and \TYPES be countable sets of node labels and relationship types, respectively. 
A property graph is a tuple \GRAPH $=$ $\langle$\NODES, \RELATIONSHIPS, \SOURCE, \TARGET, \PROPERTYFN, \LABELSFN, \TYPESFN $\rangle$ where:

\begin{itemize}
    \item \NODES is a finite subset of \NODEIDS, whose elements are referred to as the nodes of \GRAPH.
    \item \RELATIONSHIPS is a finite subset of \RELIDS, whose elements are referred to as the relationships of \GRAPH.
    \item \SOURCE : \RELATIONSHIPS $\to$ \NODES is a function that maps each relationship to its source node.
    \item \TARGET : \RELATIONSHIPS $\to$ \NODES is a function that maps each relationship to its target node.
    \item  \PROPERTYFN : (\NODES $\cup$ \RELATIONSHIPS) $\times$ \PROPERTYKEYS $\to$ \VALUES is a finite partial function that maps a (node or relationship) identifier and a property key to a value.
    \item \LABELSFN : \NODES $\to 2$\textsuperscript{\LABELS}   is a function that maps each node id to a finite (possibly empty) set of labels.
    \item \TYPESFN : \RELATIONSHIPS $\to$ \TYPES is a function that maps each relationship identifier to a relationship type.
\end{itemize}

\definition{Tables.}
Let \NAMES be a countable set of names. 
A record is a partial function from names to values, conventionally denoted as a tuple with named fields $\RECORD = (\NAME[1] : \VALUE[1], \dots , \NAME[n] : \VALUE[n])$ where $\NAME[1], \dots , \NAME[n]$ are distinct names, and $\VALUE[1], \dots , \VALUE[n]$ are values. 
The order in which the fields appear is only for notation purposes. 
We refer to $\DOMAIN(u)$, i.e., the domain of \RECORD, as the set $\lbrace\NAME[1], \dots , \NAME[n]\rbrace$ of names used in \RECORD. 
%Two records \RECORD and \RECORD['] are uniform if $\DOMAIN(\RECORD) = \DOMAIN(\RECORD['])$. If $\RECORD = (\NAME[1] : \VALUE[1], \dots , \NAME[n] : \VALUE[n])$ and $\RECORD['] = (\NAME[1]['] : \VALUE[1]['], \dots , \NAME[n]['] : \VALUE[n]['])$ are two records, then $(\RECORD,\RECORD['])$ denotes the record $(\NAME[1] : \VALUE[1], \dots , \NAME[n] : \VALUE[n], \NAME[1]['] : \VALUE[1]['], \dots , \NAME[n]['] : \VALUE[n]['])$, assuming that all $\NAME[i] , \NAME[j][']$ for $i \leq n, j \leq m$ are distinct. 

%If $\NAMES = \lbrace \NAME[1], \dots, \NAME[n] \rbrace$ is a set of names \VALUE is a value, then $(\NAMES : \VALUE)$ denotes the record $(\NAME[1] : \VALUE, \dots , \NAME[n] : \VALUE)$. 
We use $()$ to denote the empty record, i.e., the partial function from names to values whose domain is empty. 
If \NAMES is a set of names, then a table with fields \NAMES is a bag, or multiset, of records \RECORD such that $\DOMAIN(\RECORD) = \NAMES$. 
A table with no fields is just a bag of copies of the empty record. 
%A table \TABLE is a bag of records, each one having multiplicity greater than zero.
%We use the notation $\TABLE = \lbrace \BAGELEM[1]: m_1, \dots ,\BAGELEM[n]:m_n\rbrace$ to indicate a table having $n$ records, where each record $\BAGELEM[i]$ has multiplicity $m_i$ with $m_i > 0$.

%In most cases, the set of fields of tables will be clear from the context, and will not be explicitly stated.
%Given two tables \TABLE and \TABLE[]['], we use $\TABLE \uplus \TABLE[][']$ to denote their bag union, in which the multiplicity of each record is the sum of their multiplicities in \TABLE and \TABLE[]['].

%If $\BAG = \lbrace \BAGELEM[1], \dots ,\BAGELEM[n]\rbrace$ is a bag, and $\TABLE[b_1], \dots ,\TABLE[b_n]$ are tables, then $\biguplus_{\BAGELEM \in \BAG} \TABLE[b]$ stands for $\TABLE[b_1] \uplus \dots \uplus \TABLE[b_n]$.

%%%%%%%%%%%%%%%%%%%%%%%%%%%%%%%%%%%%%%%%%%%%%%%%%%%%%%%%%%%%%%%%%%%%%%%%%%%%%%%%%%%%%
%                               Evaluation semantic                                 %
%%%%%%%%%%%%%%%%%%%%%%%%%%%%%%%%%%%%%%%%%%%%%%%%%%%%%%%%%%%%%%%%%%%%%%%%%%%%%%%%%%%%%

\subsection{Query language}\label{section:backgroud:evaluation-semantic}
The \CYPHER query language includes expressions, patterns, clauses, and queries.
For brevity, we only focus on clauses and queries.
The syntax of \CYPHER query and clauses is depicted in Figure \ref{figure:cypher-sintax}.

\begin{figure}[t]
\centering
\begin{lstlisting}[frame=single, mathescape=true]
query ::= query$^\circ$ | query UNION query | query UNION ALL query
query$^\circ$ ::= RETURN ret | clause query$^\circ$
ret ::= $\textcolor{red}{*}$ | expr [AS a] | | ret $\textcolor{red}{,}$ expr [AS a]
clause ::= [OPTIONAL] MATCH pattern_tuple [WHERE expr]
            | WITH ret [WHERE expr] | UNWIND expr AS a
pattern_tuple ::= pattern | pattern$\textcolor{red}{,}$ pattern_tuple
\end{lstlisting}
\caption{Syntax of queries, and clauses of \CYPHER}
\label{figure:cypher-sintax}
\end{figure}

A query is either a sequence of clauses ending with the \code{RETURN} statement, or a union of two queries.
The semantics of queries associates a query Q and a graph G with a function $[[Q]]_G$ that takes a table and returns a table.
Note that the semantics of a query Q is a function and it should not be confused with the output of Q. 
The evaluation of a query starts with the table containing one empty tuple, which is then progressively changed by applying functions that provide the semantics of Q’s clauses. 
The composition of such functions, i.e., the semantics of Q, is a function again, which defines the output as
$$output(Q,G) = [[Q]]_G(T())$$
where T() is the table containing the single empty tuple ().

\section{Formal specification of \SERAPH}
\label{section:formal-specification}
As discussed in the introduction, the main objective of \SERAPH is to extend compositionally the semantics and the syntax of \CYPHER in order to handle streams of graphs and enable continuous query answering.
The key elements of \SERAPH are as follows:
\begin{itemize}
    \item a data model that extends the \CYPHER data model to model streams of property graphs;
    \item a continuous evaluation semantics that formally specifies the evaluation time instants and fully reuses \CYPHER semantics; and
    \item a syntax that extends compositionally the syntax of the \CYPHER query language to provide continuous query answering.
\end{itemize}

Section~\ref{section:formal:data-model} discusses the data model, while Section~\ref{section:formal:query-language} introduces the query language and defines the evaluation semantics.

\subsection{\SERAPH data model}
\label{section:formal:data-model}

In this section, we explain how the data model of \CYPHER can be extended to deal with streams of graphs.
One of the first steps towards this goal consists in considering the temporal dimension that Property Graphs in such data streams must have.
We start by defining the notion of time as in \cite{DBLP:journals/vldb/ArasuBW06}.

\definition{Time.}
The time $T$ is an infinite, discrete, ordered sequence of time instants $(t_1,t_2,\dots)$, where $t_i \in N$. 
A time unit is the difference between two consecutive time instants $(t_{i+1} - t_i)$ and it is constant.

It is now possible to extend the definition of Property Graph with a temporal annotation, and consequently define Property Graph streams as sequences of Property Graphs.

\definition{Timestamped Property Graph.}
A timestamped Property Graph is pair $(d, t)$, where $d$ is a Property Graph and $t \in T$ is a time instant. 

\definition{Property Graph Stream.}
A Property Graph Stream $S$ is a (potentially) unbounded sequence of timestamped Property Graph in non-decreasing time order:

$$S = ((d_1,t_1),(d_2,t_2),(d_3,t_3),(d_4,t_4),\dots)$$

where, for every $i > 0$, $(d_i,t_i)$ is a timestamped Property Graph and where $t_i \leq t_{i+1}$. Figures \ref{figure:example-administers}, \ref{figure:example-tested-for}, and \ref{figure:example-is-with} are examples of timestamped property graphs in a property graph stream.

We introduce now the concepts of the time-varying Property Graph and instantaneous Property graph. 
Intuitively, time-varying graphs capture the dynamic evolution of a graph over time, while instantaneous graphs represent the content of the graph at a fixed time instant.

\definition{Time-varying Property Graph.}
A time-varying graph \TVPG is a function that relates time instants $t \in T$ to Property Graph:
$$\TVPG : T \to \{g\ |\ g\ is\ a\ Property\ Graph\}$$
Given a Time-varying Property Graph \TVPG we use the term instantaneous Property Graph $\TVPG(t)$ to refer to the Property Graph identified by the time-varying graph $\TVPG$ at the given time instant $t$.

% \definition{Instantaneous Property Graph.} An instantaneous Property Graph $\TVPG(t)$ is the Property Graph identified by the time-varying graph $\TVPG$ at the given time instant $t$.

For example, Figures \ref{figure:example-window-1} and \ref{figure:example-window-2} respectively represent the Instantaneous Property Graph obtained by the union of the events occurred from 5:28 AM to 9:28 AM, and the one obtained by the union of the events from 5:36 AM to 9:36 AM.

To conclude the property graph extensions, we define two concepts that will be used when introducing the window operators: the consistency of two property graphs and, consequently, the union operation of two property graphs.
Such concepts allows for automatically merging two or more property graphs.
Consequently, the \SERAPH user does not have to care about how two or more events are merged together.

\definition{Consistency of two Property Graphs.}
Assume that \GRAPH[1] $=$ $\langle$\NODES[1], \RELATIONSHIPS[1], \SOURCE[1], \TARGET[1], \PROPERTYFN[1], \LABELSFN[1], \TYPESFN[1] $\rangle$ and \GRAPH[2] $=$ $\langle$\NODES[2], \RELATIONSHIPS[2], \SOURCE[2], \TARGET[2], \PROPERTYFN[2], \LABELSFN[2], \TYPESFN[2] $\rangle$ are Property Graphs. 
\GRAPH[1] and \GRAPH[2] are consistent if: 
\begin{enumerate}
    \item for every $r \in \RELATIONSHIPS[1] \cap \RELATIONSHIPS[2]$, it holds that  $\SOURCE[1](r) =  \SOURCE[2](r)$, $\TARGET[1](r) =  \TARGET[2](r)$ and $\TYPESFN[1](r) =  \TYPESFN[2](r)$
    
\end{enumerate}

\definition{Union of Property Graphs.}
Under the unique name assumption \cite{tao2010integrity}, we can define the union of two Property Graphs.
If \GRAPH[1] and \GRAPH[2] are not consistent, then $\GRAPH[1] \cup \GRAPH[2]$ is defined as the empty Property Graph. Otherwise, $\GRAPH[1] \cup \GRAPH[2] = (\NODES[1] \cup \NODES[2], \RELATIONSHIPS[1] \cup \RELATIONSHIPS[2], \SOURCE, \TARGET, \PROPERTYFN, \LABELSFN, \TYPESFN)$, where:
\begin{enumerate}
    \item for every $x \in (\NODES[1] \cup \NODES[2])$: $\LABELSFN(x)=\LABELSFN[1](x) \cup  \LABELSFN[2](x)$
    \item for every $x \in (\NODES[1] \cup \NODES[2] \cup \RELATIONSHIPS[1] \cup \RELATIONSHIPS[2])$ and $k \in \PROPERTYKEYS: \PROPERTYFN(x,k)=\PROPERTYFN[1](x,k) \cup  \PROPERTYFN[2](x,k)$
\end{enumerate}

The match clause, as defined in the \CYPHER specification, can operate over an instantaneous property graph. Generalising, in our model each operator processes instantaneous inputs and produces instantaneous outputs; the sequence of instantaneous inputs (outputs) at different time instants are time-varying inputs (outputs). 
It follows that we need to define the time-varying and instantaneous extensions for tables, too.

\definition{Time-varying Tables.}
% In the previous section, we introduced the notions of time-varying and instantaneous property graphs, to take into account the time dimension.
% The time-varying property graph \TVPG is a mapping between the time and the set of property graphs, and given a time instant $t$ the instantaneous graph $\TVPG(t)$ identifies an instantaneous property graph.
A time-varying table \TVT maps time instants $t \in T$ to the set of tables:

$$\TVT: T \to \lbrace \psi | \psi~is~a~table\rbrace$$

Given a time-varying table \TVT, we use the term instantaneous table $\TVT(t)$ to refer to the table identified by the time-varying table at the given time instant $t$.

Lastly, we need to define the bag difference of two tables. Given two tables \TABLE and \TABLE[]['], we use $\TABLE \smallsetminus \TABLE[][']$ to denote their bag difference, in which the multiplicity of each record is the difference of their multiplicities in \TABLE and \TABLE[][']. For example, Table \ref{table:example-output-1} and \ref{table:example-output-2} represent the Instantaneous Table resulting from the bag difference of the complete results of the query respectively at 9:28 AM and at 9:36 AM. Indeed, Chase does not appear in Table \ref{table:example-output-2} because it appeared in Table \ref{table:example-output-1}.

\subsection{\SERAPH query language}
\label{section:formal:query-language}
In the previous section, we defined the \SERAPH data model by adding the temporal dimension in the \CYPHER model in three different ways: Timestamped Property Graphs are property graphs with a time annotation; Property Graph Streams are ordered sequences of Timestamped Property Graphs, and Time-varying and Instantaneous Property Graphs capture the changes of a Property Graph over time. 
In this section, we present \SERAPH, an extension of the \CYPHER language to continuously query this data model.
One of the main differences between \SERAPH and \CYPHER is the way in which queries are evaluated.
\CYPHER allows to issue one-time queries, queries that are evaluated once by the \CYPHER engine. 
In contrast, \SERAPH allows to register continuous queries, queries issued once and continuously evaluated.
Such queries are evaluated multiple times, and the answer is composed by streaming out the results of each evaluation iteration or, often, the bag difference of the most recent evaluation and the previous one.

We present the definition of a \SERAPH query, which extends compositionally the notion of \CYPHER query presented in the Section~\ref{section:backgroud}.

\definition{\SERAPH query.} A \SERAPH query associates a query SQ and a Time-varying Property Graph \TVPG with a function $[[SQ]]_{\TVPG}$ that takes a table and a time instant and returns a table.
We refer to the sequence of evaluation time instants as ET.

In the remaining of this section we formalize the extensions mentioned in Section~\ref{section:example}, namely the \textit{Window Operators}, the \textit{Stream Operators}, and the \textit{Continuous evaluation semantic}.

\subsubsection{Window operators}
We introduce the concept of window over a Property Graph Stream, which create a Property Graph by extracting relevant portions of the Property Graph Stream (e.g., Line \ref{lst:seraph_query_window} in Figure \ref{figure:example-seraph-query}). 

\definition{Window.}
%A window $W(S)$ is a set of Property Graphs extracted from a stream $S$. 
The output of a window $W(S)$ is a Property Graph resulting from the union of a set of Property Graphs contained in the Property Graph Stream.
$$W(S) = \cup \lbrace d | d \in set(S) \rbrace$$

Time-based windows and event-based windows select a set of the Property Graphs contained in the Property Graph Stream.

\definition{Time-based window.}
A time-based window \TBW is a window defined through two time instants $\OPENT, \CLOSET$ (respectively named opening and closing time instants) such that:
$$\TBW(S) = \cup \lbrace d | (d, t) \in S \land t \in (\OPENT, \CLOSET] \rbrace$$

In order to be able to process the content of the stream at different time instants, we need an operator that creates multiple property graphs over the stream (i.e. a time-varying property graph).

\definition{Time-based sliding window operator.}
A time-based sliding window operator \TBSW is defined through the width parameter \WIDTH and takes as input a stream S and produces a time-varying Property Graph $\GRAPH[\TBSW]$. 

At each time instant \WTIME, $\GRAPH[\TBSW](t)$ contains the content of the time-based window $\TBW = (\WTIME[] - \WIDTH, \WTIME[]]$, i.e., 
$$\TBSW(S, t) = \GRAPH[\TBSW](t) = \lbrace d | (d, t) \in S \land t \in (\WTIME[] - \WIDTH, \WTIME[]] \rbrace$$

\definition{Event-based window.}
A time-based window \TBW is a window defined through one time instant $\CLOSET$, and a number $N$ representing the number of Property Graphs to be extracted from the Property Graph Stream.
Intuitively event-based window defines its output by extracting the last $N$ events of S with the largest timestamps $\leq \CLOSET$ (or all events if the length of S up to $\CLOSET$ is $\leq  N$).

In order to be able to process the content of the stream as new event arrives, we need an operator that creates a time-varying property graph.

\definition{Event-based sliding window operator.}
An event-based sliding window operator is defined thorough the parameter $N$, and takes as input a stream S and produces a time-varying Property Graph graph $\GRAPH[\TBSW]$.
Intuitively, an event-based sliding window defines its output over time by sliding an event-based window over the ordered stream S. 

\subsubsection{Streaming operators}
The streaming operators require a time instant as input parameter because they are time-aware.
They need to know the current evaluation time in order to produce their outputs.
They reintroduce the temporal dimension in the data, appending a time instant on the tables.
The streaming operators can be considered the dual operators of window operators, that process timestamped property graph removing the time annotation.
Those operators were first defined for relational data stream processing in \cite{DBLP:journals/vldb/ArasuBW06}. 
For this reason, we maintain the original names and we redefine them to work in the \SERAPH setting. For the syntax, we refer the readers to the \code{EMIT} clause in Section \ref{section:formal-specification:syntax-and-semantic}.

We start by defining the RStream operator.

\definition{RStream operator.}
Let \TVT be a time-varying table and $t$ the evaluation time instant. 
We define $RStream$ in the following way:
$$RStream(\TVT, t) = \lbrace(\mu, t)|\mu \in \TVT(t)\rbrace$$

The RStream operator is the simplest one among the three that we present in this section.
It takes as input a time-varying table and annotates the instantaneous table with the evaluation time t. 
This operator allows streaming out the whole answer produced at each evaluation iteration.

\definition{IStream operator.} 
Given a time-varying table \TVT, and two consecutive time instants $t_{j-1}$ and $t_j$ in the sequence of evaluation time instants, we define the $IStream$ operator as follows:

$$IStream(\TVT, t_j, t_{j-1}) = \lbrace(\mu, t_j)~|~\mu \in \TVT(t_j) \smallsetminus \TVT(t_{j-1})\rbrace$$

IStream streams out the difference between the answer of the current evaluation and the one of the previous iteration. 
IStream generally produces shorter answers and it is used in cases where it is important to put the focus on what is new.

\definition{DStream operator.} 
Given a time-varying table \TVT, and two consecutive time instants $t_{j-1}$ and $t_j$ in the sequence of evaluation time instants, we define the $DStream$ operator as follows:
$$DStream(\TVT, t_j, t_{j-1}) = \lbrace(\mu, t_j)~|~\mu \in \TVT(t_{j-1}) \smallsetminus \TVT(t_j)\rbrace$$

The output produced by DStream is the part of the answer at the previous iteration that is not in the current one.

\subsubsection{Syntax and evaluation semantic}
\label{section:formal-specification:syntax-and-semantic}

Starting from the grammar shown in the background section, we define the syntax of a \SERAPH query as  in Figure~\ref{figure:seraph-sintax}.

\begin{figure}
\centering
\begin{lstlisting}[frame=single, mathescape=true]
query$^\sim$ ::= REGISTER QUERY <id> {
              FROM STREAM <source>
              STARTING FROM time_instant
              WITH WINDOW RANGE range 
              query 
              [ CONSTRUCT 
                    CREATE pattern_tuple 
                RETURN GRAPH ]
              EMIT streaming_operator EVERY range
              INTO <destination>
           }
streaming_operator ::= ON ENTERING | ON EXIT | SNAPSHOT
range ::= event_range | <ISO_8601_duration>
event_range ::= <integer> Events | 1 Event
time_instant ::= Earliest | Latest | <ISO_8601_datetime>
\end{lstlisting}
\caption{Syntax of \SERAPH queries.}
\label{figure:seraph-sintax}
\end{figure}

The \code{REGISTER QUERY} clause allows for registering a new query into the \SERAPH system.
The \code{FROM STREAM} clause allows for specifying the source of the Property Graph Stream.
The \code{STARTING FROM} clause defines the first evaluation time instant.
The width parameter of sliding windows is defined using the \code{WINDOW RANGE} clause.
Inspired by \textit{morpheus}\footnote{https://github.com/opencypher/morpheus}, the optional \code{CONSTRUCT CREATE} \texttt{pattern\_tuple} \code{RETURN GRAPH} clause allows for creating a Property Graph Stream. 
The \code{EMIT} clause determines which streaming operator is used.
In particular, the \code{SNAPSHOT} clause specifies that the $RStream$ operator has to be used, while the \code{ON ENTERING} and \code{ON EXIT} clauses allow for selecting $IStream$ and $DStream$, respectively.
Using the \code{EVERY} clause, together with the \code{STARTING FROM} clause, the sequence of evaluation time instance can be determined.
In particular, the \code{EVERY} clause defines the frequency of the evaluation.
Such a frequency can be specified either with an ISO 8601 duration or in terms of number of events.
The \code{STARTING FROM} clause, instead, defines the 
first evaluation time instant as an ISO 8601 datetime.
Alternatively, the keyword \code{Earliest} (\code{Latest}) defines the first evaluation time instant as the datetime associated with the first (last) event.
Finally, the \code{INTO} clause allows for specifying the destination of the result stream.

%In particular, the starting time instant $t_0$ can be defined thorough the \code{STARTING FROM} clause.

At this point, \SERAPH operators can process instantaneous inputs and produce instantaneous outputs.
What we need to do now is to model the continuous evaluation process. 
To do it, we include the evaluation time in the \CYPHER evaluation semantics.
Then, we explain that the continuous query answering is done by executing the query at each time instant of the sequence ET (the evaluation time instants defined in the \SERAPH query presented at the beginning of the section). 

The semantics of queries associates a query SQ and a graph Time-varying Property Graph \TVPG with a function $[[SQ]]_{\TVPG}$ that takes a table and a time instant and returns a table.
The evaluation of a query starts with the table containing one empty tuple, which is then progressively changed by applying functions that provide the semantics of SQ’s clauses. 
The composition of such functions, i.e., the semantics of SQ, is a function again, which defines the output as:

$$output(SQ,\TVPG,t) = [[SQ]]_{\TVPG}(T(),t)$$

This new concept requires a revision of the definitions of the existing \CYPHER evaluation of queries, clauses and expressions.
For the sake of brevity, we show the continuous evaluation semantics of \code{MATCH} clause.

$$[[\code{MATCH}~\pi]]_{\TVPG}(T, t) = [[\code{MATCH}~\pi]]_{\TVPG(t)}(T)$$

\definition{Evaluation time instants}
We defined ET as the sequence of time instants at which the evaluation occurs. 
It is an abstract concept which is key to the \SERAPH query model and its continuous-evaluation semantics, but it is hard to use it in practice when designing the \SERAPH syntax.
In fact, the ET sequence is potentially infinite, so the syntax needs a compact representation of this set.
To be noticed that the sequence does not depend on the stream processing engine, since the syntax of \SERAPH allows for defining the initial time instant $t_0$.
Still, the stream processing engine is in charge of counting the number of events $N_t$ processed after the time instant $t_0$.
The \code{EVERY} clause allows for specifying the frequency of reporting that, together with the first evaluation time instant $t_0$ defined through the \code{STARTING FROM} clause, defines the ET sequence.
In particular we can distinguish two main cases: the first one, based on time, and the other, based on events.

\textit{Time-based evaluation.}
The \code{EVERY}~\texttt{ISO\_8601\_duration} clause allows for specifying a time interval.
We refer to such time interval as $i$.
Given $i$ and $t_0$, we define the ET sequence as follows:
$$ ET = \lbrace t~|~(t - t_0) / i = 0 \rbrace$$

\textit{Event-based evaluation.}
The \code{EVERY}~\texttt{event\_range} clause allows for specifying a number of event.
We refer to such number of event as $N$.
Given $N$, and $N_t$, we define the ET sequence as follows:

$$ ET = \lbrace t~|~N_t~mod~N = 0\rbrace$$

For every $t \in ET$, a query evaluation is triggered, updating the Time-Varying Property graph resulting from the sliding window, executing the \texttt{query} and, eventually, producing a timestamped table as a result.

\section{System prototype}
\label{section:system-prototype}

In this section, we describe our reasoning and efforts towards a working prototype\footnote{https://github.com/riccardotommasini/rsp4j/tree/gsp4j} for evaluating \SERAPH queries and present a series of experiments to assess the performances of such a prototype.

\subsection{System architecture.}
Requirements R1 and R2 demand to expose direct control to the part of the data system that controls the response to input (Tick) and the result reporting (Report), directly impacting the data system design. 
To this extent, we decided to adopt RSP4J~\cite{DBLP:conf/esws/0001BOV21} for an API for fast prototyping stream processing engines, which already includes essential windowing operation and primitives that allow customizing the system's reporting.

Although RSP4J includes some graph operations (e.g., Triple Pattern Matching), the following challenges concern the realization of a graph stream processor for \SERAPH: 
\begin{inparaenum}[(i)]
    \item The Data Model: RSP4J is designed for RDF Stream Processing and does not directly support Property Graph operations.
    \item The Data Format: Despite Property Graphs' popularity as a data model, a (de-facto) standard data format is still missing. 
    \item Indexing: native graph databases make extensive use of indexing to speed up query answering performance. Neo4j, for instance, uses $BTree$ for indexing the values of any given property. \footnote{https://neo4j.com/docs/operations-manual/current/performance/index-configuration/}. 
    However, when it comes to streaming data, the frequency of ingestion prevents the effective use of indexes due to the high cost of maintenance.
    Last but not least, 
    \item to guarantee data integrity, Neo4j supports the ACID properties.  
    However, transactional behavior opposes the strict latency requirements of stream processing applications.
\end{inparaenum}

\begin{figure}[t]
    \centering
    \includegraphics[width=\linewidth]{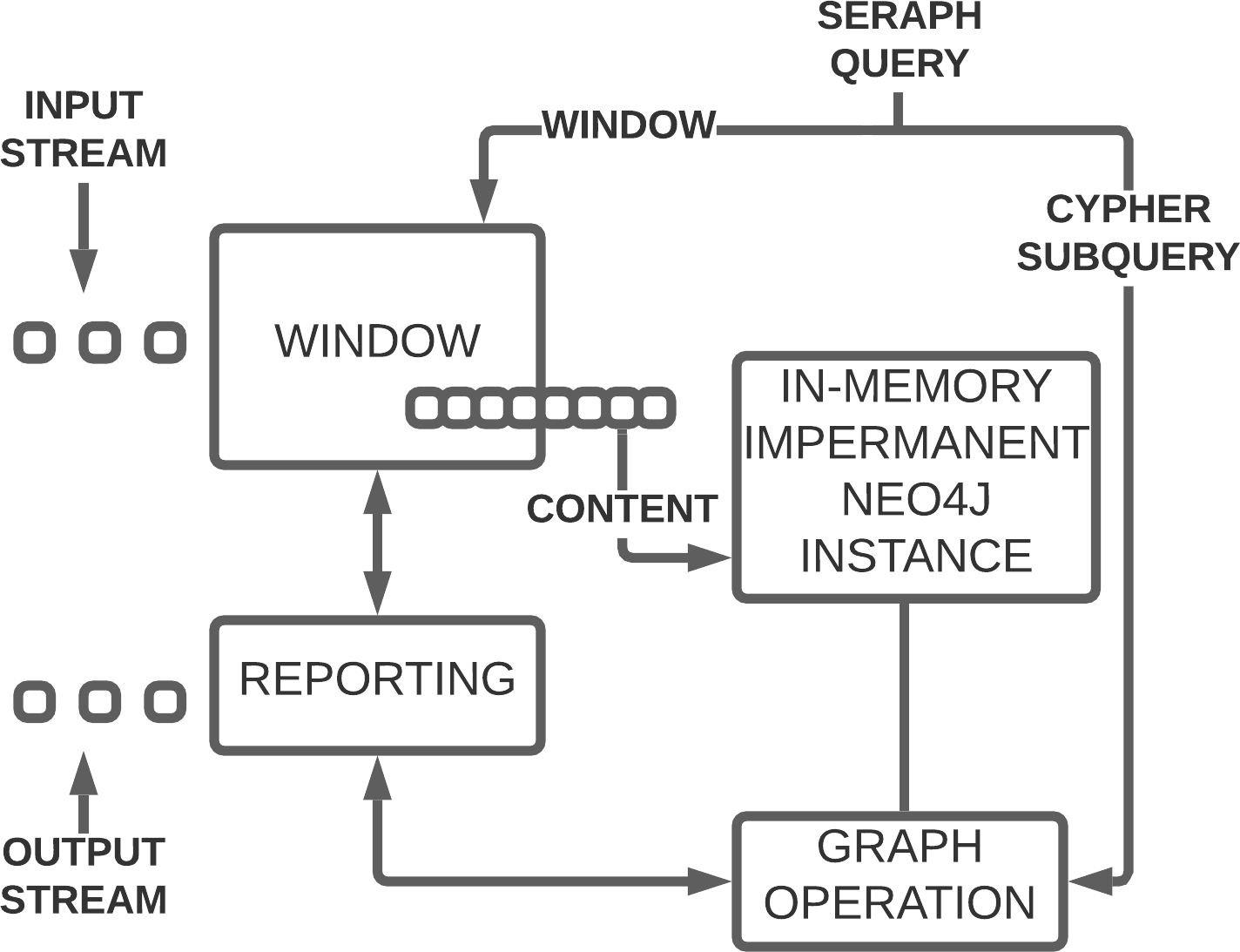}
    \caption{Seraph Engine Architectures}
    \label{fig:arch}
\end{figure}

Figure~\ref{fig:arch} presents the architecture of our system prototype. 
The system reuses RSP4J window operators, which we designed along with CQL time-based sliding windows. 
Note that the time window size directly increases the amount of data that the system has to store.
Consequently, the more significant is the amount of data for each window, the longer the query execution will take.
The reporting policy is customized to support Seraph's \texttt{EMIT} clause, which controls how to emit the output. 

As per data format, we opted for JSON-PG \cite{DBLP:journals/corr/abs-1907-03936} since we need to be graph-native at the ingestion level. 
Indeed, most of the existing graph stream processing approaches adopt a node-stream and edge-stream solution. At the same time, \SERAPH's uniqueness lies in retaining a "think like a graph" perspective within the stream.
Furthermore, JSON-PG format can describe property graph data used in existing graph databases such as Neo4j, Oracle Labs PGX, and Amazon Neptune.
Figure~\ref{figure:jsonpg} shows an example of JSON-PG serialization format, where we use a Property Graph to describe a wind speed observation performed by a specific sensor.

%\begin{lstlisting}[frame=single, escapechar=|, caption={JSON PG Example}, label={lst:jsnpg}, language=json]
%{
%  "nodes":[
%    {"id":101, "labels":["Person"], 
% x                  "country":["United States"]}}
% , {"id":102, "labels":["Person", "Student"], 
%     "properties":{"name":["Bob"], 
%                   "country":["Japan"]}}
%  ],
%  "edges":[
%    {"from":101, "to":102, "undirected":true, 
%     "labels":["likes"], 
%     "properties":{"since":[2012]}}
%  ]
%}
%\end{lstlisting}

\begin{figure}
    \centering
\begin{lstlisting}[frame=single, escapechar=|, language=json]
{
   "nodes":[
      {
         "id":2564832,
         "labels":[ "Observation" ],
         "properties":{
            "property":[ "WindSpeed" ],
            "value":[ 12.0 ],
            "unit_of_measure":[ "milesPerHour" ]
         }
      },
      {
         "id":4684,
         "labels":[ "Sensor" ],
         "properties":{
            "name":[ "FOR" ]
         }
      }
   ],
   "edges":[
      {
         "id":2564833,
         "from":4684,
         "to":2564832,
         "labels":[ "HAS_OBSERVED" ]
      }
   ]
}
\end{lstlisting}
    \caption{JSON PG Example.}
    \label{figure:jsonpg}
\end{figure}

For the query execution, we adapted Neo4J \textit{impermanent}  databases (IDB), which provide an in-memory Neo4J instance with limited transactional support. 
Neo4J IDB is not durable. 
As per window maintenance, we operate on a materialized view maintained within the Neo4J IDB. 
We use lightweight transactions (not durable) to control the windowing lifecycle, i.e., ingestion, retrieval, and purging. 

Last but not least, \SERAPH Engine directly supports consumption from Kafka.
The motivation for adopting a durable ingestion layer is two-fold.
First, we currently do not provide any fault-tolerance guarantee. 
Second, the poll-based consumption approach simplifies the performance measurement since the engine is currently synchronous.

\subsection{Performance Evaluation}
We designed a series of tests to evaluate the performance of the prototype.
During each test, the \SERAPH system consumes events from Kafka as fast as possible while filtering and executing the \CYPHER query according to the specifications.
We aim at observing the average consumption rate achievable and the average memory usage of the system.
In particular, we want to investigate how such values change w.r.t the window size.

\textbf{Data source.}
As a preliminary step, we transformed SRBench~\cite{DBLP:conf/semweb/ZhangDCC12}, a collection of meteorological data observed during known hurricanes, in JSON-PG format.
Then, a Kafka cluster, composed of a single broker running on a \texttt{Amazon EC2 t3.xlarge} instance, has been deployed to store the events relative to the Charley hurricane.
The Charley event series counts \~1.8 million Timestamped Property Graphs, each one composed of several nodes and relationships.

\textbf{Experiment description.}
Figure~\ref{figure:evaluation-seraph-query} reports the query used for our experiments.
The \code{REGISTER QUERY} clause at Line \ref{lst:eval_query_register} allows for registering the query into the system that manages \SERAPH queries with id \texttt{srbench\_charley\_observation\_count}.
At Line \ref{lst:eval_query_from_stream} the \code{FROM STREAM} clause is used to specify \texttt{kafka://sr-bench-charley} as  source stream.
The \code{STARTING FROM} clause specifies the initial time instant that is used to define the sequence of the evaluation time instants.
In this case, \code{Earliest} means that the first evaluation time instant is related to the oldest event in the source stream.
The \code{WINDOW RANGE}~\texttt{PT5M} operator at Line \ref{lst:eval_query_window} sets up a 5 minutes wide time-based sliding window.
Line \ref{lst:eval_cypher_start} and Line \ref{lst:eval_cypher_end} compute the number of observation.
Line \ref{lst:eval_query_emit} specifies the \code{EMIT} clause.
In particular, the \code{SNAPSHOT} operator directly emits the results without performing any additional operation.
The \code{EVERY} operator specifies the frequency of the evaluation process. 
Specifically, the code \code{EVERY}~\texttt{PT5M}, in combination with \code{WINDOW RANGE PT5M} sets up a time based \textit{tumbling} window. 
At last, the \code{INTO} clause is used to specify \texttt{kafka://observation-count} as the target Kafka topic.

\textbf{Execution environment.}
The system is running on an \texttt{Amazon EC2 t3.2xlarge} instance which, besides executing the continuous query, collects metrics about the consumption rate and memory usage.
In particular, the \SERAPH system runs in a Docker container with limited CPUs (2) and RAM (16 GBytes), while the metric collection task is achieved by Telegraf that monitors the Docker environment and pushes the metrics into InfluxDB.

\begin{figure}[t]
    \centering
    \includegraphics[width=\linewidth]{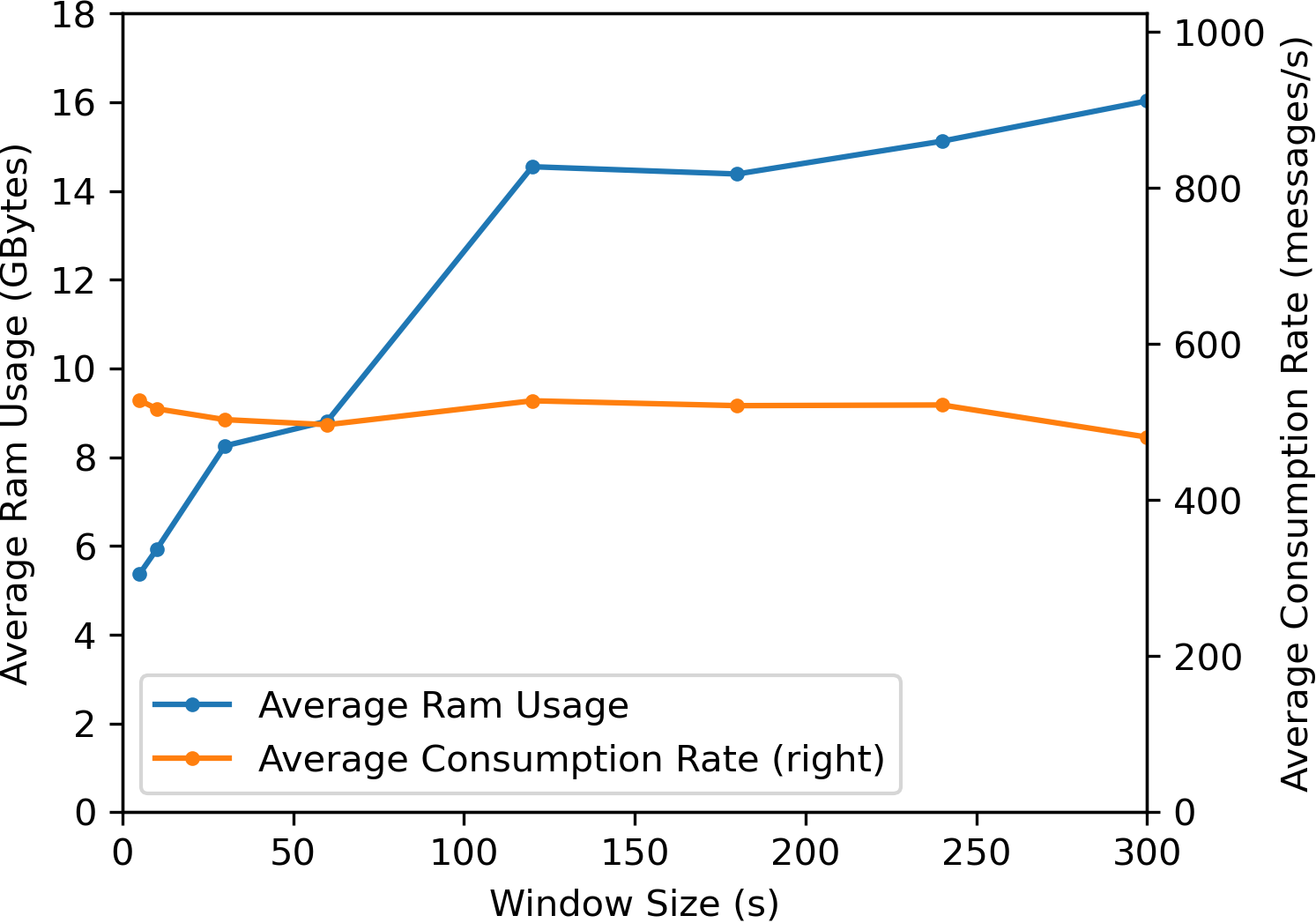}
    
    \caption{Seraph Engine Performances}
    \label{fig:performances}
\end{figure}

\textbf{Discussion.}
Figure~\ref{fig:performances} illustrates how the average memory usage and the average consumption rate change by varying the window size.
In particular, we can observe that the query execution time does not impact the average consumption rate.
However, when dealing with larger windows, the systems require more memory to store all the data belonging to the current window, thus resulting in much more intense memory usage.

\begin{figure}[t]
\centering
\begin{lstlisting}[frame=single, escapechar=^]
REGISTER QUERY srbench_charley_observation_count    ^\label{lst:eval_query_register}^
  FROM STREAM kafka://sr-bench-charley              ^\label{lst:eval_query_from_stream}^
  STARTING FROM Earliest                            ^\label{lst:eval_query_starting}^
  WITH WINDOW RANGE PT5M                            ^\label{lst:eval_query_window}^
  MATCH (o:Observation)                             ^\label{lst:eval_cypher_start}^
  RETURN COUNT(o) as observation_count              ^\label{lst:eval_cypher_end}^
  EMIT SNAPSHOT EVERY PT5M                          ^\label{lst:eval_query_emit}^
  INTO kafka://observation-count                    ^\label{lst:eval_query_into}^
}
\end{lstlisting}
\caption{\SERAPH query to count the number of observations.}
\label{figure:evaluation-seraph-query}
\end{figure}

\section{Industrial use cases}
\label{section:industrial-use-cases}

In this section we present two use cases that showcase how \SERAPH can simplify the encoding of continuous property graph queries in a network monitoring and in a crime investigation scenarios.

\subsection{Network Monitoring}

Computer networks span all levels of the stack from physical connections up to mobile and web-applications connecting networks of users. It is well known that Graph Databases offer a natural way of modelling, and storing all these types of computer networks\footnote{A number of commercial Network and IT Management solutions exploit Graph Databases and there are also open source solutions like Mercator (\url{https://github.com/LendingClub/mercator}) and the Assimilation Project (\url{http://assimilationsystems.com/}).}.

While graph query languages like \CYPHER play a key role in investigating dependencies and in running diagnostic analyses (e.g., root cause of a past network fault), \SERAPH offers the possibility to continuously execute real-time impact analysis of network events.

Let's assume that we model the network endpoints (e.g., servers, routers, switch, interfaces, and racks) of the data center as nodes and the "cables" between them as relationships. For instance, a rack \texttt{HOLDS} a switch that \texttt{ROUTES} an interface that \texttt{CONNECTS} a router in a network zone. 

We want to continuously monitor the network connectivity in our datacenter watching for anomalous routes. Let's assume that the connections are redundant, i.e., if one of our cables gets loose or cut, i.e. the \texttt{ROUTES} relationship between a switch’s interface and the network breaks, the number of hops can increase, but no rack can become unreachable. We know from the configuration of the network that the shortest routes from all racks to the egress router\footnote{An egress router is a Label Switch Router that is an end point (drain) for a given Label Switched Path.} requires on average 5 hops, but network events may cause this path to be longer and we observed a standard deviation of 0.3 hops. We can identify anomalous routes using the z-score\footnote{In statistics, the z-score is the number of standard deviations $\sigma$ by which an individual $x$ is above or below the mean value $\mu$ of the population. It is calculated as $(x-\mu)/\sigma$} and watching for routes whose length has a z-score larger than 3, i.e., it is longer than 99,9\% of the paths. 

\begin{figure}
    \centering
\begin{lstlisting}[frame=single, escapechar=|]
REGISTER QUERY watch_for_anomalous_routes {         |\label{lst:nm_seraph_query_01}|
    FROM STREAM kafka://network-monitoring          |\label{lst:nm_seraph_query_02}|
    STARTING FROM Latest                            |\label{lst:nm_seraph_query_03}|
    WITH WINDOW RANGE PT10M                          |\label{lst:nm_seraph_query_04}|
    MATCH path = allShortestPaths(                  |\label{lst:nm_seraph_query_05}|
            (rack:Rack)-[:HOLDS|:ROUTES|:CONNECTS*]-(router:Router:Egress))          |\label{lst:nm_seraph_query_06}|
    WITH rack, avg(length(path)) as 10minAvg, path         |\label{lst:nm_seraph_query_07}|
    WHERE (10minAvg - 5 / 0.5) >= 3             |\label{lst:nm_seraph_query_08}|
    RETURN path                                     |\label{lst:nm_seraph_query_09}|
    EMIT SNAPSHOT EVERY PT1M                        |\label{lst:nm_seraph_query_10}|
    INTO kafka://anomalous-routes                   |\label{lst:nm_seraph_query_11}|
}
\end{lstlisting}
    \caption{Using \SERAPH to watch for paths in a computer network that are anomalously long.}
    \label{figure:network-management-seraph}
\end{figure}

Figure \ref{figure:network-management-seraph} illustrates how to encode this need in a \SERAPH query. The source of the streaming graphs is a Kafka topic specified at  Line \ref{lst:nm_seraph_query_02}, i.e., \code{kafka://network-monitoring}. We assume that it is populated by a Kafka stream application that creates a time-varying graphs capturing the dynamic evolution of the computer network end-points and of the "cables" among them. In this way, at each time instant, an instantaneous graph represent the configuration of the entire network.
The query uses the \code{WITH WINDOW} (Line \ref{lst:nm_seraph_query_04}) and the \code{EMIT ... EVERY} (Line \ref{lst:nm_seraph_query_10}) clauses to define a 10 minutes wide \emph{hopping window} that reports every minute (i.e., \code{PT1M}). The query finds the shortest paths from each rack to the egress router (Lines \ref{lst:nm_seraph_query_05} and \ref{lst:nm_seraph_query_06}) and computes the average length of those paths in the last 10 minutes (Line \ref{lst:nm_seraph_query_08}). If the z-score of those paths related to each rack (Line \ref{lst:nm_seraph_query_07}) is greater than 3, then it writes the anomalous path (Line \ref{lst:nm_seraph_query_09}) in the Kafka topic \code{kafka://anomalous-routes} (Line \ref{lst:nm_seraph_query_11}).

\subsection{Crime Investigations}

\begin{figure}[h!]
    \centering
    \includegraphics[width=0.7\linewidth]{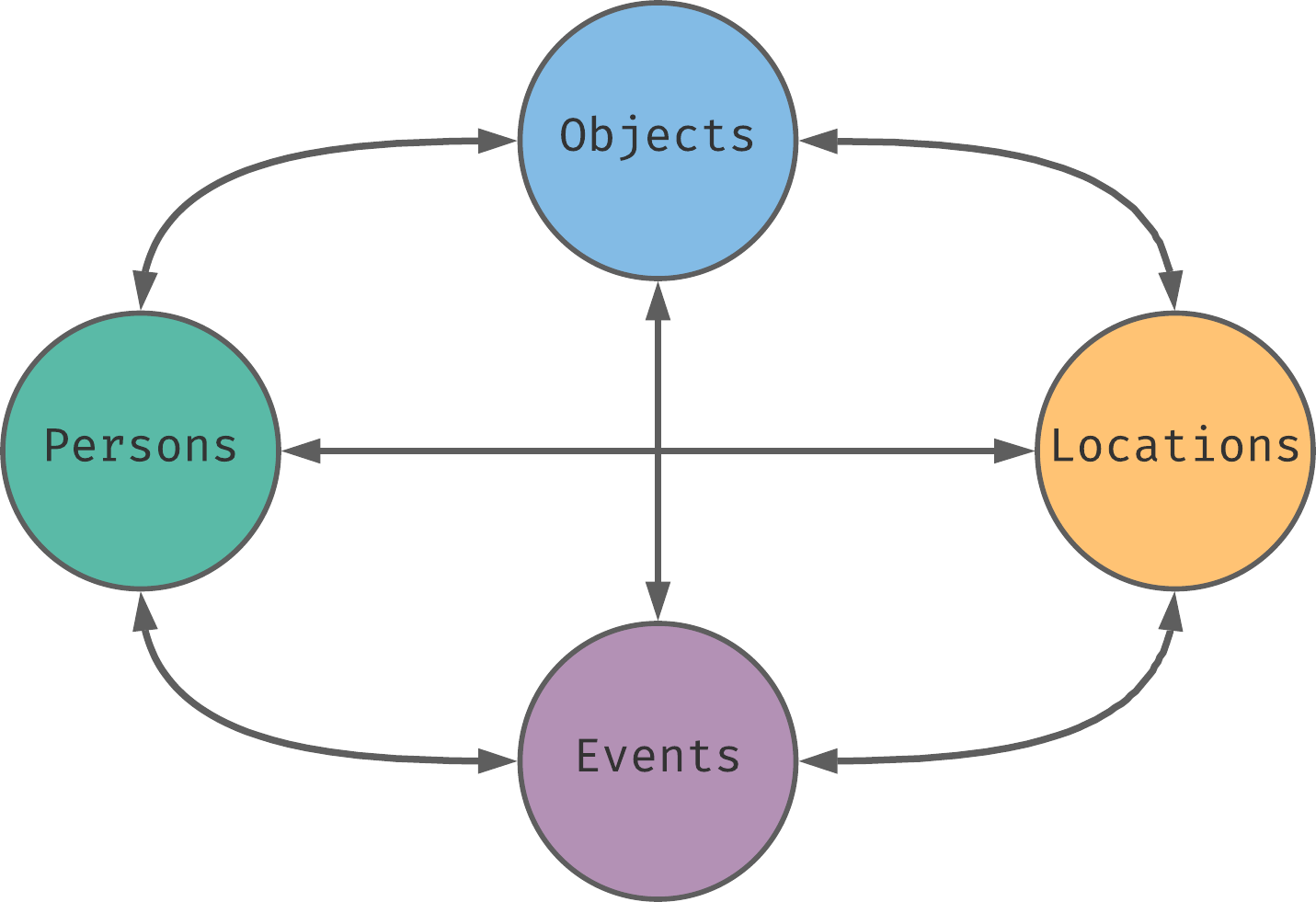}
    \caption{The Person-Object-Location-Event data model.}
    \label{fig:pole-seraph}
\end{figure}

From fraud detection, to security, encompassing surveillance and contact-tracing,
investigations often require \textit{connecting the dots}. Data models like POLE (Person-Object-Location-Events) (cf. Figure~\ref{fig:pole-seraph}) underpin a number of analyses that require the identification of patterns\footnote{\url{https://www.techuk.org/component/techuksecurity/security/download/2302?file=Breaking_down_barriers_Oct_2014_FINAL.pdf}}. POLE was originally intended for historical analyses that one can perform using graph query languages like \CYPHER.
However, POLE is event-centric thus it already includes temporal metadata that \SERAPH can exploit unlocking a number of additional analyses including, but not limited to, real-time surveillance and reactive contact-tracing.

Assuming we are adopting the POLE model for surveillance, we model \textit{crimes} and \texttt{calls} as \texttt{Event}s, which  \texttt{OCCURRED\_AT} a \texttt{Location}. Moreover, we assume that a number of smart-cameras, which can identify each \texttt{Person} passing by (cf. \texttt{NEAR\_TO}), are deployed in different \texttt{Location}s within the city of \textit{London}. 
We are interested in monitoring \textit{who} is passing by crime-scenes and detecting potential suspects. In this example, we consider suspects whoever has been convicted (cf. \texttt{PARTY\_TO}) for a crime of the same type of the one reported.
Moreover, considering that on average a person walks about 5km in an hour\footnote{\url{https://en.wikipedia.org/wiki/Walking}}, we restrict the scope of the monitoring to an area of 3km from the crime scenes and a time range of 15 minutes.
Like in the networking use-case, we assume the input streams are maintained by a Kafka stream application and accessible as Kafka topics. Moreover, we assume to have access to the database of known criminals and their felonies. 

Figure \ref{fig:pole-seraph} illustrates how to encode the information-need above in a \SERAPH query. The source and the sink of the streaming query are respectively the Kafka topics \code{kafka://scotland-yard} (Line \ref{lst:seraph:crime:source}) and \code{kafka://suspects} (Line \ref{lst:seraph:crime:sink}). The query focuses on the last 15 minutes, reporting every 5 minutes. To this extent, it uses the \code{WITH WINDOW} clause (Line \ref{lst:seraph:crime:window}) and controls the results reporting using the \code{EMIT ... EVERY} (Line \ref{lst:seraph:crime:emit}). The query monitors the streams of crime reports (Lines~\ref{lst:seraph:crime:match1-a}-\ref{lst:seraph:crime:match1-b}) and crosschecks if anyone, who is identified by a smart-camera, was a convicted criminal (Lines~\ref{lst:seraph:crime:match2-a}-\ref{lst:seraph:crime:match2-b}). To restrict the search space, the query looks only for cameras within 3000 meters from the crime scenes and to those suspects that had taken part in a crime of the same type before.
The functions \texttt{point} and \texttt{distance} allows \SERAPH users to perform geo-spatial comparisons. As an output, the query returns the last seen location, the suspect description, and the crime references (Line~\ref{lst:seraph:crime:return}). %\todo{why 3km and 15min}

\begin{figure}
\begin{lstlisting}[frame=single, escapechar=|]
REGISTER QUERY watch_for_suspects {  |\label{lst:seraph:crime:qname}|
   FROM STREAM kafka://scotland-yard/calls |\label{lst:seraph:crime:source}|
   STARTING FROM Latest |\label{lst:seraph:crime:starting-from}|
   WITH WINDOW RANGE PT15M |\label{lst:seraph:crime:window}|
   MATCH (call:Event)--[:OCCURRED_AT]->(l1:Location), |\label{lst:seraph:crime:match1-a}|
    WITH new, point(l1) AS crime_scene |\label{lst:seraph:crime:match1-b}|
   MATCH (crime:Event)<-[:PARTY_TO]-(person:Suspect)-[:NEAR_TO]->(last_seen:Location) |\label{lst:seraph:crime:match2-a}|
    WITH person, last_seen, distance(point(last_seen), crime_scene) AS distance |\label{lst:seraph:crime:match2-b}|
   WHERE distance < 3000 |\label{lst:seraph:crime:where}| AND new.type=old.type
   RETURN person, last_seen, call.description AS crime_description |\label{lst:seraph:crime:return}|
   EMIT SNAPSHOT EVERY PT5M |\label{lst:seraph:crime:emit}|
   INTO kafka://suspects |\label{lst:seraph:crime:sink}|
}
\end{lstlisting}
\caption{Using \SERAPH to look for suspects around crime scenes.}
\label{fig:seraph:crime}
\end{figure}

\section{Related Work}
\label{section:related-work}
% A lot of effort has been put in formalizing the extension of existing languages to handle data streams.
In this section, we discuss the work that directly relates to the proposed extensions. In particular, we position \SERAPH in the state of the art and we discuss its relation with other work in the area of graphs and stream processing.
% In particular, we discudescribe various graph query languages and briefly discuss how the new GQL\footnote{https://www.gqlstandards.org/} standard solves the interoperability challenges.
% Then, we survey the stream extension of various query languages, discussing the common aspects.
% Finally, a description of a generic Stream Processing Engine is reported, highlighting the fact that a query language is only one of the building blocks of a complete system.

\emphasis{Dynamic Graphs and Streaming Graphs} are two additional attempts to extend existing graph data models with a temporal dimension~\cite{DBLP:journals/corr/abs-1912-12740}. \textit{Dynamic Graphs} are graphs whose content, i.e., vertices and edges, is unpredictably updated. Updates take the form of insertions and deletions. The data system that manages the dynamic graph either stores the most recent version of the graph or the graph's entire change history. Differently from \SERAPH, languages for querying dynamic graphs do not necessarily require continuous semantics, but scalable and time-sensitive query answering in presence of changes is one of the key research focuses. \textit{Streaming Graphs} are dynamic graphs that grow indefinitely. Updates are typically limited to insertions and, as for \SERAPH, query answering must take unboundedness into account. Moreover, the data management system is assumed to be unable to store the whole graph state, therefore it focuses on the finite sub-graph that is relevant for the query answering. Although this approach is similar to \SERAPH's windowing, it does not rely on time- or order-aware window operators but instead tries to compute approximated answer. 

\begin{figure}[h]
    \begin{center}
    \includegraphics[width=7cm]{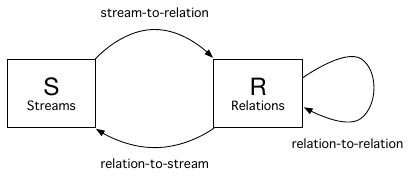}
    \caption{CQL: operator classes and mappings used in abstract semantics} \label{CQL}
    \end{center}
\end{figure}

\emphasis{Declarative Stream Processing Languages} have been around for two decades. Most of the existing solutions, including those associated with the Big Data initiative~\cite{DBLP:journals/sigmod/HirzelBBVSV18}, present an SQL-like syntax~\cite{DBLP:conf/edbt/0001SVJ20} and build upon the Continuous Query Language model (CQL)~\cite{DBLP:journals/jucs/Buza06}.

CQL prescribes to make the management of (relational) streams orthogonal to the management of relations. 
Figure \ref{CQL} shows how the three families of operators, which CQL defines, relate to each other.
Stream-to-Relation (S2R) operators produce a relation from a stream and are based on the concept of a sliding window over a stream, i.e., a window that contains a finite portion of the stream. Relation-to-Relation (R2R) operators produce a relation from one or more other relations and are derived from traditional relational queries expressed in SQL. Relation-to-Stream (R2S) operators produce a stream from a relation. 

\SERAPH follows CQL orthogonalisation principle because it makes the language compositional and  maintainable~\cite{DBLP:journals/sigmod/Date84}. 
However, differently from CQL, \SERAPH provides the primitives to fully control the reporting. Learning from Dindar et al.~\cite{DBLP:journals/vldb/DindarTMHB13}, who showed how the operational semantics of Stream Processing engines is often uncontrollable by the user, \SERAPH's \code{EMIT} clause gives an end-to-end view of what impacts execution semantics from inputs to output. \SERAPH users have full control on the query execution semantics. 
The query language that \SERAPH resembles the most is the RDF Stream Processing Query Language (RSP-QL)~\cite{dell2014rsp}. RSP-QL was proposed by the Semantic Web community in the late 2000s' to accommodate the need for processing heterogeneous data streams. RSP-QL extends CQL work on RDF Streams. Dell'Aglio et al. introduced new families of operators based on CQL's S2R and R2S, as well as SPARQL 1.1 algebra \cite{sparql}. Such operator families allow writing continuous SPARQL queries on RDF streams.
\SERAPH differs from RSP-QL in its focus on streams of Property Graphs and on its goal to compositionally extent \CYPHER. 

\section{Future work}
\label{section:future-work}
This is preliminary work.
Further extensions are needed in order to have a query language that is able to accommodate more use-cases.

\textbf{Static graph.} 
Support for multiple property graphs is required to combine static knowledge with streams of events. Considering our running example, this feature would allow combining a static property graph that describes the organizations of shops across the city, with the stream of events.
Moreover, the real-world of \textit{Network Monitoring} (cf Section~\ref{section:industrial-use-cases}) shows a clear need for distinguishing between a  \textit{static} graph source representing the network configuration and the stream of events happening on the network.
Multiple graph support is already available with \CYPHER 10\footnote{https://github.com/opencypher/openCypher} and Morpheus\footnote{https://github.com/opencypher/morpheus}.
However, to the best of our knowledge, a formal specification of \CYPHER 10 is still missing. As an alternative, GQL is supposed to support multi-graphs query. Thus, future GQL formalization can e exploited to achieve such a goal.

\textbf{Multi-stream.}
Multi-stream support allows developers to perform queries across streams. Considering our running example in Section~\ref{section:example}, we can imagine to distinguish  a stream describing when the coupons are used, limiting the notifications only to the users that have not used their coupons yet.
In real-world scenarios like those in Section~\ref{section:industrial-use-cases}, the need for processing data from multiple data sources is common. In particular, the \textit{Crime Investigation} use-case highlights the need for consuming distinguished streams, e.g., a stream of crime reports and a stream of cameras observations. This will allow defining different windows operators tailored to tame the velocity of each stream separately.

\textbf{Complex Event Recognition (CER).}
Stream processing languages are not limited to continuous analytics. Indeed, solutions for CER typically allow for filtering, combining, and transforming events~\cite{DBLP:journals/vldb/GiatrakosAADG20,artikis_et_al:DR:2020:13058}. The most important features in CER languages are temporal operators that allow detecting patterns of  events over unbounded streams. Extending \SERAPH for complex event recognition will benefit a number of use-cases that require complex stream transformations. For instance, in the context of the \textit{Crime Investigation} use-case, we can imagine to monitor the movements of suspects upon the occurrence of a crime that may involve the suspects.

\section{Conclusions}
\label{section:conclusions}
In this paper, we presented \SERAPH, a graph query language that compositionally extends the semantics and the syntax of \CYPHER 9 for dealing with streams of property graphs and continuous query answering.
We introduced a running example that showcases the capabilities of the introduced language and, in particular, the continuous evaluation paradigm that allows for viewing the continuous evaluation as a sequence of instantaneous evaluations.
\SERAPH extends \CYPHER formally specifying:
\begin{enumerate*}[label=\roman*)]
  \item the \SERAPH data model that extends the data model of \CYPHER to handle streams of graphs;
  \item the continuous evaluation semantics of \SERAPH that, based on \CYPHER semantics, offers the possibility to merge events under unique name assumption automatically, specifies operators to transform streams of graphs in graphs, frames the concept of sequence of evaluation time instants, and defines operators to transform back the result of the evaluations in a stream of graphs; and
  \item the \SERAPH syntax that extends the one of \CYPHER to provide continuous query answering.
\end{enumerate*}
The present work aims to lays the keystone for future continuous graph query languages, such as the continuous extension of GQL (the new ISO standard for graph query languages).
Even if, in this paper, we show that \SERAPH is exploitable in industrial use cases, some improvements are still needed.
In particular, we foresee three important extensions:
\begin{enumerate*}[label=\roman*)]
  \item multi-stream support that allows for consuming two or more streams at a time,
  \item multi-window support, i.e., the ability to define multiple sliding windows in the same query, and
  \item static graph support that allows for exploiting existing knowledge base graphs.
\end{enumerate*}

\balance
%\input{content/acknowledgments}

%\clearpage
\bibliographystyle{ACM-Reference-Format}
\bibliography{biblio}

\end{document}